\documentclass[preprint,10pt]{elsarticle}
\usepackage{lipsum}
\makeatletter
\def\ps@pprintTitle{
 \let\@oddhead\@empty
 \let\@evenhead\@empty
 \def\@oddfoot{}
 \let\@evenfoot\@oddfoot}
 \renewcommand{\MaketitleBox}{
  \resetTitleCounters
  \def\baselinestretch{1}
  \begin{center}
    \def\baselinestretch{1}
    \Large \@title \par
    \vskip 18pt
    \normalsize\elsauthors \par
    \vskip 10pt
    \footnotesize \itshape \elsaddress \par
  \end{center}
  \vskip 12pt
}

\usepackage{amssymb}
\usepackage{amsthm}
\usepackage{lineno}
\usepackage{amsfonts}
\usepackage{amsmath}
\usepackage{graphicx}
\usepackage{cases}
\usepackage{epstopdf}
\usepackage{caption}
\usepackage{algorithmic}
\usepackage{mathabx}
\biboptions{sort&compress}

\ifpdf
  \DeclareGraphicsExtensions{.eps,.pdf,.png,.jpg}
\else
  \DeclareGraphicsExtensions{.eps}
\fi

\usepackage{amssymb}
\usepackage{mathtools}

\usepackage[mathscr]{eucal}
\usepackage{color}
\usepackage{geometry}
\usepackage{tikz}
\usetikzlibrary{shapes,arrows}
\usetikzlibrary{decorations.pathreplacing,angles,quotes}

\usepackage{todonotes}

\usepackage{dsfont}
\usepackage{amsthm}
\usepackage{thmtools}
\usepackage{cleveref}
\usepackage{verbatim}
\usepackage{graphicx}
\usepackage{subfig}
\usepackage{tikz}
\usepackage{pgfplots}

\newtheorem{theorem}{Theorem}[section]
\newtheorem{lemma}[theorem]{Lemma}
\newtheorem{proposition}[theorem]{Proposition}

\numberwithin{equation}{section}

\newtheorem*{example}{Example}
\newtheorem{remark}{Remark}
\newtheorem{assumption}{A\!\!}[section]

\begin{document}

\begin{frontmatter}
\title{Mean Field Games of Optimal Trading with an SPDE Model of the Limit Order Book 
}
\author[1]{Hanchao Liu}  \author[2]{Dena Firoozi\corref{cor1}}
\address[1]{Department of Decision Sciences, HEC Montréal, Montréal, QC, Canada\\
(email: hanchao.liu@hec.ca)}
\address[2]{Department of Statistical Sciences, University of Toronto, Toronto, ON, Canada\\ (email: dena.firoozi@utoronto.ca)}
\cortext[cor1]{Corresponding author.}
\begin{abstract}
We develop a mean field game (MFG) model for optimal trading in which the interaction among traders is mediated by the dynamics of the limit order book. The bid- and ask-side order densities are modeled as stochastic processes taking values in infinite-dimensional function spaces, and solve semilinear stochastic partial differential equations (SPDEs) with stochastic, time-varying order submission rates and multiplicative infinite-dimensional noise. This extends an existing limit order book framework to capture the aggregate impact of strategic traders executing market orders on the order densities. In this way, trading activity affects the order densities, which determine the market depths near the best quotes and hence the mid-price, and the mid-price, in turn, affects the traders’ decisions. We establish the well-posedness and regularity of the resulting density-price system and use it to formulate an $N$-player optimal trading game in which the traders’ market orders enter the book through their aggregate order flow. We then study the corresponding MFG, which involves semilinear SPDEs driven by common noise processes and a highly nonlinear stochastic differential equation (SDE) for the mid-price, as the number of traders tends to infinity. Under regular market conditions, we characterize the mean field (the limiting aggregate trading rate) as a fixed point of a nonlinear consistency condition, expressed as a forward-backward SPDE-SDE system, and prove its existence and uniqueness under suitable technical conditions. Finally, we show that the limiting optimal trading strategy induces an $\epsilon_N$-Nash equilibrium for the $N$-player game, with $\epsilon_N = O(N^{-1/2})$, and provide a numerical illustration of the mean field consistency equations. 

\end{abstract}
\end{frontmatter}

\section{Introduction}
Optimal trading and execution\footnote{In this paper, the terms optimal trading and optimal execution are used interchangeably.} concern the execution of a trade at the lowest possible cost. The central difficulty is that trading is not costless, since purchases tend to exert upward pressure on a security’s price while sales tend to exert downward pressure, so that the act of trading moves the market adversely against the trader. Consequently, submitting a large order in a single transaction can be prohibitively costly, which motivates traders to divide it into smaller trades executed over time. However, slow execution entails its own cost, since the longer a trade remains incomplete, the greater the trader’s exposure to adverse price movements. Optimal strategies therefore balance these two competing effects, trading gradually enough to limit the trader’s own price impact, yet rapidly enough to curtail exposure to market risk.

Single-agent optimal trading problems have been studied extensively; see, e.g., \cite{JaimungalBook2015, Sebastian2013, Almgren2001, Alfonsi2010}. In these problems, a single trader aims to liquidate or acquire a given number of shares over a fixed time horizon, choosing an optimal trading rate. Several extensions of this basic setting have been considered. For instance, \cite{Bayraktar2011} studies a partially observed setting in which the market liquidity is not observed, \cite{Forde2022, AbiJaber2025} incorporate price signals and transient price impact, and \cite{Cartea2023} accounts for latency in the marketplace. In these formulations, the trader's own activity generates price impact, whereas the strategic activity of the other market participants is either ignored or enters through exogenous processes, such as price signals or order flow, that do not react to the trader's strategy. 

In practice, however, prices move as a result of the collective behaviour of many strategic traders who continually adjust their strategies in response to changing market conditions. A dynamic game model of optimal trading is proposed in \cite{CarmonaLacker2015}, in which a large number of traders compete with each other and affect prices through their average trading rate. This model was introduced as an example problem amenable to the mean field game (MFG) methodology. 

MFG theory provides a framework for dynamic games with a large number of interacting agents, each of whom has a negligible individual impact on the population as its size grows, while the interaction among them is captured through an aggregate quantity, such as the empirical mean or distribution of the agents' states or controls, whose limit as the number of agents tends to infinity is called the mean field. The resulting limiting game is a tractable approximation of the original finite-player game and, under suitable conditions, yields strategies that form an approximate Nash equilibrium when the population is sufficiently large \cite{huang2006large,huang2007large,lasry2007mean,carmona2018probabilistic,bensoussan2013mean}.

MFG models have found natural applications in financial markets, where many agents interact through common market variables, including systemic risk \cite{carmona2013mean,Bo2015SystemicRiskInterbanking,chang2022Systemic}, portfolio trading \cite{Cardaliaguet2018Meanfieldgame,Fu-Horst-2021,Fu-Horst-2022}, equilibrium pricing \cite{Firoozi2022MAFI,gomes_mean-field_2018,fujii_mean_2021,firoozi_shrivats_jaimungal_2022_principalagentmeanfield}, and electricity markets \cite{Rene-Aid-2021,alasseur2023}. Optimal trading and execution is an important class of such applications. When many traders are present, their collective activity affects market conditions and hence the trading problem faced by each of them. This gives rise to a mean field interaction in which each trader's optimal strategy depends on the aggregate order flow, which is in turn generated by the strategies of all traders. The goal of the MFG formulation is to capture the collective impact of the traders on prices and to characterize a market equilibrium, i.e., a set of strategies from which no trader has an incentive to deviate (see e.g. \cite{FirooziISDG2017,Cardaliaguet2018Meanfieldgame,huang2019mean,firoozi2019mean,Fu-Horst-2021,casgrain_meanfield_2020}). Related to this line of research, $N$-player game formulations of optimal execution problems are studied for instance in \cite{Nutz-Campbell-2026-a,Nutz-Campbell-2026-b}.

However, to the best of our knowledge, in existing dynamic game models of optimal trading, the interaction among traders enters through a reduced-form price impact mechanism, in which the aggregate trading rate affects the asset price directly. While tractable, this formulation does not describe the mechanism through which trading activity changes the state of the limit order book and, in turn, the asset price. In a limit order book market, market orders consume the liquidity available on the ask- and bid-side sides, so the evolution of the book is a natural channel through which aggregate trading activity affects prices. 

The purpose of this paper is to develop an MFG model for optimal trading in which the interaction among traders is mediated through the dynamics of the limit order book. The dynamics of the limit order book have been studied extensively in the literature. In particular, partial differential equation (PDE) models of the limit order book are considered, for instance, in \cite{lasry2007mean, MarkowichTeichmannWolfram2016, HorstKreher2017, Horst-Paulsen-2017}, while stochastic partial differential equation (SPDE) models are studied in, e.g. \cite{HorstKreher2019, HamblyKalsiNewbury2020, cont2021}. We model the ask- and bid-side densities as stochastic processes taking values in infinite-dimensional function spaces, whose dynamics are described by SPDEs whose drift depends on the aggregate trading rate of market participants, extending the infinite-dimensional limit order book model of \cite{cont2021}. The corresponding market depths are obtained from the order densities near the best bid and ask, and the mid-price dynamics then follow from the evolution of these depths, adapting the approach of \cite{cont2021}. Consequently, aggregate trading activity affects the order densities, changes in the order densities affect the market depth and the mid-price, and the resulting mid-price in turn affects the optimal trading decisions of the agents. This creates a feedback mechanism between the individual trading behaviour of the agents and the dynamics of the limit order book as a whole.

The main contributions of the paper are as follows.
\begin{itemize}
    \item We extend the limit order book SPDE model of \cite{cont2021} to a broader class of order densities, which, in particular, allows us to incorporate dynamic interactions among traders. The extension has two main features. First, in contrast to the constant order submission rates in \cite{cont2021}, we allow these rates to evolve as stochastic processes. Second, the fluctuations in order flow are driven by an infinite-dimensional multiplicative noise, which allows different price levels to be driven by different sources of randomness, whereas \cite{cont2021} uses a one-dimensional Brownian motion on each side of the book. We then study the well-posedness of the resulting SPDEs and derive regularity properties of the solution that are required for our subsequent analysis, particularly for obtaining a stochastic differential equation (SDE) for the mid-price.

\item Using the results above, we incorporate the aggregate trading flow of traders submitting market orders into the evolution of the ask- and bid-side order densities, each modeled by a semilinear SPDE, and derive the resulting mid-price dynamics in the form of a nonlinear SDE. This yields a coupled density-price system in which price impact arises endogenously from the SPDE dynamics of the limit order book, rather than being specified in reduced form directly through the mid-price dynamics, as is common in the existing literature. It also provides a framework for an $N$-player game of optimal trading, in which each trader chooses a strategy to minimize execution costs while facing the price impact generated by the collective trading activity of all traders.

\item We then study the corresponding MFG, which involves semilinear SPDEs driven by common noise processes and a highly nonlinear mid-price SDE, as the number of traders $N$ tends to infinity. Under regular market conditions, we characterize the mean field (the limiting aggregate trading rate) as a fixed point of a nonlinear consistency condition, expressed as a forward-backward SPDE-SDE system, and establish its existence and uniqueness. Finally, we show that the limiting optimal trading strategy induces an $\epsilon_N$-Nash equilibrium for the $N$-player game, with $\epsilon_N = O(N^{-1/2})$.
\end{itemize}

The remainder of this paper is organized as follows. Section~\ref{lobd} introduces the stochastic limit order book model and establishes the density-price dynamics, together with the required well-posedness and regularity results. Section~\ref{NPgm} connects the trading activity of the individual agents to the dynamics of the limit order book and formulates the $N$-player optimal trading game. Section~\ref{FMGg} derives the corresponding MFG, characterizes the mean field equilibrium, establishes its existence and uniqueness, and provides a numerical illustration of the equilibrium dynamics. Finally, Section~\ref{sec:epsilon-nash} shows that the mean field equilibrium is an $\epsilon$-Nash equilibrium for the finite-player game.

\section{Limit Order Book Dynamics} \label{lobd}

In this section, we introduce the dynamics of the limit order book. We model
the evolution of the ask- and bid-side order densities, from which the corresponding
market depths and mid-price dynamics are determined. Our formulation generalizes
the stochastic limit order book model of \cite{cont2021} to a broader class of densities, which subsequently allows us to incorporate dynamic interactions among the traders participating in the market. We also establish
additional regularity results for the order density processes, which allow
certain quantities derived from the order book to be expressed as diffusion
processes. These results will be used in the subsequent analysis.
\subsection{The Density-Price System}\label{dpsys}
In this section, we introduce the stochastic limit order book model that
determines the evolution of the ask- and bid-side order densities around the
mid-price and the corresponding mid-price dynamics. The density processes are modeled by
stochastic partial differential equations (SPDEs), while the mid-price
dynamics are determined by the evolution of liquidity on the two sides of
the book. 
This density-price system will subsequently be coupled with the optimal
execution problem of the traders to form an $N$-player game.
Throughout this paper, we consider a complete probability space
$(\Omega,\mathcal F,\mathbb P)$. Let $H$ be a Hilbert space and let
$\mathcal G=(\mathcal G_t)_{t\in[0,T]}$ be a filtration. We set
$\Omega_T:=[0,T]\times\Omega$, and denote by $\mathcal G_P$ the progressive
$\sigma$-field on $\Omega_T$ associated with $\mathcal G$. We denote by 
\begin{equation}
\mathcal M_{\mathcal G}(0,T;H)
=
\left\{
X:[0,T]\times\Omega\to H:
X \text{ is } \mathcal G_T\text{-measurable and }
\mathbb E\int_0^T\|X_t\|_H^2\,dt<\infty
\right\}.
\end{equation}
When $H=\mathbb R$, we simply write $\mathcal M_{\mathcal G}(0,T)$.

We define the Hilbert spaces
\begin{equation}
H^a:=L^2(0,L),
\qquad
H^b:=L^2(-L,0),
\qquad L>0.
\end{equation}
Let $W^a$ and $W^b$ be $Q_a$- and $Q_b$-Wiener processes on
$H^a$ and $H^b$, respectively, where $Q_a$ and $Q_b$ are positive
trace-class covariance operators. We set
\begin{equation}
H_0^a:=Q_a^{1/2}(H^a),
\qquad
H_0^b:=Q_b^{1/2}(H^b).
\end{equation} For the ask- and bid-side limit order book
densities, defined on the filtered probability space
$(\Omega,\mathcal F,(\mathcal F_t)_{t\in[0,T]},\mathbb P)$, we consider the SPDEs given by 
\begin{equation}
\label{eq:general_lob}
\begin{cases}
du_t^a
=
\left[
A_au_t^a+V^a(t,\omega,u_t^a)
\right]dt
+
B^a(u_t^a)\,dW_t^a,
\\[2mm]
du_t^b
=
\left[
A_b u_t^b+V^b(t,\omega,u_t^b)
\right]dt
+
B^b(u_t^b)\,dW_t^b,
\\[2mm]
u_0^a\in H^a,
\qquad
u_0^b\in H^b,
\end{cases}
\end{equation}
where $W^a$ and $W^b$ are adapted to $(\mathcal F_t)_{t\in[0,T]}$. The operators \(A_a\) and \(A_b \) are given by
\begin{align}
A_au
&=
\eta_a\frac{d^2u}{dx^2}
+
\beta_a\frac{du}{dx}
+
\alpha_a u,
\qquad x\in(0,L),
\\
A_b u
&=
\eta_b\frac{d^2u}{dx^2}
-
\beta_b\frac{du}{dx}
+
\alpha_b u,
\qquad x\in(-L,0),
\end{align}
where
\begin{equation}
\eta_a,\eta_b>0,
\qquad
\beta_a,\beta_b>0,
\qquad
\alpha_a,\alpha_b<0.
\end{equation}
The operators \(A_a\) and \(A_b \) are endowed with homogeneous
Dirichlet boundary conditions. More precisely,
\begin{align}
D(A_a)
&=
\left\{
u\in H^2(0,L):
u(0)=u(L)=0
\right\},
\\
D(A_b )
&=
\left\{
u\in H^2(-L,0):
u(-L)=u(0)=0
\right\},
\end{align}
where $H^2$ denotes the Sobolev space of order two. Both $A_a$ and $A_b $ generate analytic semigroups,
denoted by $(S_a(t))_{t\geq 0}$ and $(S_b(t))_{t\geq 0}$,
respectively. Moreover, for every $T>0$, both semigroups are uniformly bounded on $[0,T]$, i.e.,
\begin{equation}
\sup_{0\leq t\leq T}\|S_l(t)\|_{\mathcal L(H^l)}
\leq M,
\qquad l\in\{a,b\},
\end{equation}
for some constant $M>0$.

The mappings
\begin{equation}
V^a:\Omega_T\times H^a\longrightarrow H^a,
\qquad
V^b:\Omega_T\times H^b\longrightarrow H^b
\end{equation}
represent the general drift terms of the ask- and bid-side limit order book
dynamics, respectively.

Similarly, the noise coefficients are mappings
\begin{equation}
B^a:H^a\longrightarrow L_2(H_0^a,H^a),
\qquad
B^b:H^b\longrightarrow L_2(H_0^b,H^b),
\end{equation}
where $L_2(H_0^l,H^l)$ denotes the space of Hilbert--Schmidt operators from
$H_0^l$ to $H^l$, for $l\in\{a,b\}$.

The processes $u^a$ and $u^b$ represent the densities of limit orders on the ask and bid sides of the limit order book, respectively, that is, the volume of outstanding limit orders per unit of relative price. More precisely, $u_t^a(x)$, $x\in(0,L)$, is the density of sell limit orders at time $t$ and relative price coordinate $x$, defined as the distance above the current mid-price, while $u_t^b(x)$, $x\in(-L,0)$, is the density of buy limit orders at distance $|x|$ below it. The volume of orders in a relative price range $[x_1,x_2]$ is thus given by the integral of the density over that range.
 The operator $A_\ell$, for $\ell\in\{a,b\}$, describes the redistribution and removal of outstanding limit orders on the corresponding side of the order book. In particular, the second-order spatial derivative represents order modifications in which existing orders are canceled and symmetrically replaced at nearby price levels, leading to a diffusion of order density across the book. The first-order derivative captures the systematic relocation of orders toward the mid-price, while the zeroth-order term accounts for cancellations that reduce the outstanding order volume without replacement. The terms $V^a$ and $V^b$ represent additional changes in the ask- and bid-side order book densities, respectively, that are not captured by the operators $A^a$ and $A^b$, such as those resulting from the arrival of new limit orders or market orders.

The broader class of order densities considered here extends the limit order book SPDE model of \cite{cont2021} in two main ways. First, unlike the constant order submission rates $f^a$ and $f^b$ in \cite{cont2021}, our model allows $V^a$ and $V^b$ to evolve as stochastic processes, capturing the time-varying and stochastic nature of order arrivals. Second, we model the stochastic terms, which capture fluctuations in order flow (e.g., from uninformed traders), by an infinite-dimensional, $H^\ell$-valued multiplicative noise. This allows different price levels to be driven by different sources of randomness, whereas \cite{cont2021} models the noise by a one-dimensional Brownian motion on each side of the order book. 

Following the price formation mechanism proposed in \cite{cont2021,cont2014price}, we relate the dynamics of the mid-price to the order flow and the available liquidity at the top of the limit order book. Let $D_t^a$ and $D_t^b$ denote the market depths on the ask and bid sides, respectively. In relative price coordinates, representing distance from the mid-price, these quantities are defined by the volumes of limit orders over a small neighborhood of the top of the book (mid-price) on each side. In particular, for a sufficiently small $\delta>0$ representing the tick size,
\begin{equation}
D_t^a
=
\int_0^\delta u_t^a(x)\,dx,
\qquad
D_t^b
=
-\int_{-\delta}^0 u_t^b(x)\,dx.
\end{equation}

Following \cite{cont2021,cont2014price}, we define the mid-price process $F$ through the relative changes in the market depths given by
\begin{equation}
\label{eq:mid_price}
dF_t
=
\kappa
\left(
\frac{dD_t^b}{D_t^b}
-
\frac{dD_t^a}{D_t^a}
\right),
\end{equation}
where $\kappa>0$ is a price-impact coefficient. In the next section, we show  that the above definition is mathematically well-defined under suitable conditions.
\subsection{Well-Posedness and Regularity of the Ask and Bid Order Density SPDEs}
We now study the solution and regularity of the SPDEs introduced in Section~\ref{dpsys}. Since the analyses of the SPDEs on the negative and positive half-intervals are analogous, we study only the ask side, and drop the superscript $a$ for simplicity. Namely,  
\begin{equation}
\label{eq:spde_positive_side}
\begin{cases}
du_t
=
\left[
Au_t+V(t,\omega,u_t)
\right]dt
+
B(u_t)\,dW_t,
\\[2mm]
u_t(0)=u_t(L)=0,
\\[2mm]
u_0\in H,
\end{cases}
\end{equation}
where
\begin{equation}
H=L^2(0,L),
\qquad
Au
=
\eta\frac{d^2u}{dx^2}
+
\beta\frac{du}{dx}
+
\alpha u.
\end{equation}
Compared with the formulation in \cite{cont2021}, the above SPDE allows for
a broader class of drift terms through the mapping $V$ and is driven by
infinite-dimensional noise. We also derive stronger regularity properties
of the solution that are necessary for our subsequent analysis as explained below.

A process $u\in \mathcal{M}_{\mathcal F}(0,T;H)$ is called a mild solution of \eqref{eq:spde_positive_side} if, for every $t\in[0,T]$,
\begin{equation}
u_t
=
S(t)u_0
+
\int_0^t S(t-s)V(s,\omega,u_s)\,ds
+
\int_0^t S(t-s)B(u_s)\,dW_s,
\qquad \mathbb{P}-\text{a.s.}
\end{equation}
Equivalently, $u$ is also a weak solution such that, for every $h\in D(A^*)$, where $A^*$ denotes the adjoint of $A$, and $t\in[0,T]$,
\begin{equation}
\langle u_t,h\rangle_H
=
\langle u_0,h\rangle_H
+
\int_0^t
\left[
\langle u_s,A^*h\rangle_H
+
\langle V(s,\omega,u_s),h\rangle_H
\right]ds
+
\left\langle
\int_0^t B(u_s)\,dW_s,h
\right\rangle_H,
\qquad \mathbb{P}-\text{a.s.}
\end{equation}
Under standard measurability, Lipschitz continuity, and linear growth assumptions on $V$, the existence and uniqueness of a solution $u\in \mathcal{M}_\mathcal F(0,T;H)$ is a well-established result in the literature; see, for example, \cite{da2014stochastic}. However, this level of regularity is not sufficient for the price dynamics considered above. Indeed, the market depth is defined by
\begin{equation}
D_t
=
\int_0^\delta u_t(x)\,dx
=
\left\langle u_t,\mathbf{1}_{[0,\delta]}\right\rangle_H,
\end{equation}
where $\mathbf{1}_{[0,\delta]}$ denotes the indicator function of the interval $[0,\delta]$, that is, $\mathbf{1}_{[0,\delta]}(x)=1$ for $x\in[0,\delta]$ and $\mathbf{1}_{[0,\delta]}(x)=0$ for $x\in(\delta,L]$. For a general solution $u\in \mathcal{M}_\mathcal F(0,T;H)$, the process $D$ is not necessarily an It\^o diffusion. Consequently, the differential $dD_t$ need not admit the usual stochastic differential equation (SDE) representation, and hence the price dynamics in \eqref{eq:mid_price} is not, in general, mathematically well-defined at this level of regularity. Moreover, without an SDE representation for the price process, it would be difficult to incorporate the resulting price dynamics into a standard stochastic control or dynamic game framework. To address this regularity issue, we introduce the following interpolation space associated with $A$, borrowing techniques from \cite{da2014stochastic,lunardi1995analytic}. For $0<\alpha<\frac12$, define
\begin{equation}
D_A\left(\alpha+\frac12,2\right)
=
\left\{
x\in H:
\int_0^\infty
t^{-2\alpha}
\|AS(t)x\|_H^2\,dt
<
\infty
\right\}.
\end{equation}

The space $D_A\left(\alpha+\frac12,2\right)$ is endowed with the norm
\begin{equation}
\|x\|_{D_A\left(\alpha+\frac12,2\right)}
=
\left(
\int_0^\infty
t^{-2\alpha}
\|AS(t)x\|_H^2\,dt
\right)^{1/2},
\end{equation}
under which it is a separable Hilbert space given $A$ is negative type, i.e., $\|S(t)\|\le Me^{-c t}$ for some $M\ge 1$ and $c>0$. We first note that the operators
\begin{equation}
A
=
\eta\frac{d^2}{dx^2}
+
\beta\frac{d}{dx}
+
\alpha,
\qquad
A^*
=
\eta\frac{d^2}{dx^2}
-
\beta\frac{d}{dx}
+
\alpha,
\end{equation}
have eigenfunctions
\begin{equation} \label{eignh}
h_n(x)
=
\sqrt{\frac{2}{L}}\,
e^{-\frac{\beta}{2\eta}x}
\sin\left(\frac{n\pi x}{L}\right),
\qquad
h_n^*(x)
=
\sqrt{\frac{2}{L}}\,
e^{\frac{\beta}{2\eta}x}
\sin\left(\frac{n\pi x}{L}\right).
\end{equation}
with the corresponding eigenvalues
\begin{equation}
\lambda_n
=
\alpha-\frac{\beta^2}{4\eta}
-\eta\left(\frac{n\pi}{L}\right)^2.
\end{equation}
The families $\{h_n\}_{n\geq1}$ and $\{h_n^*\}_{n\geq1}$ form a pair of biorthogonal Riesz bases of $H$ since these families are obtained from the standard sine basis under the bounded invertible multiplication operators $ e^{\pm\frac{\beta}{2\eta}x}$.
Consequently, every $x\in H$ admits the unique expansion
\begin{equation}
x
=
\sum_{n=1}^{\infty}
\langle x,h_n^*\rangle_H h_n,
\end{equation}
and
\begin{equation}\label{rzbound}
d_1\|x\|_H^2
\leq
\sum_{n=1}^{\infty}
\left|\langle x,h_n^*\rangle_H\right|^2
\leq
d_2\|x\|_H^2,
\end{equation}
where $d_1,d_2>0$ are constants independent of $x$. We assume that $\lambda_1<0$, so that both $A$ and $A^*$ are of negative type. This allows us to characterize the regularity induced by
$D_A(\theta,2)$ in terms of the Riesz basis introduced above. The following result will be useful in the sequel.

By convention, for any Banach space $E$, we denote its dual space by $E'$. Moreover, throughout the remainder of the paper, the constant $C>0$ may vary from one place to another.

\begin{proposition}\label{Dfunctional}
Let $\frac34<\theta<1$. For every $x\in D_A(\theta,2)$, the series
\begin{equation}
\Lambda(x)=\sum_{n=1}^{\infty}
\left(
\int_0^\delta h_n(\xi)\,d\xi
\right)
\langle x,A^*h_n^*\rangle_H,
\end{equation}
converges absolutely. Hence, the series defines an element $\Lambda\in \bigl(D_A(\theta,2)\bigr)'$.
\end{proposition}

\begin{proof}
Since $A^*h_n^*=\lambda_n h_n^*$, we have $\langle x,A^*h_n^*\rangle_H=\lambda_n\langle x,h_n^*\rangle_H$. Therefore, 
\begin{equation}
\sum_{n=1}^{\infty}
\left|
\left(\int_0^\delta h_n(\xi)\,d\xi\right)
\langle x,A^*h_n^*\rangle_H
\right|
\leq\sum_{n=1}^{\infty}
\left|\int_0^\delta h_n(\xi)\,d\xi\right|
|\lambda_n|\,
|\langle x,h_n^*\rangle_H| .
\end{equation}
The explicit form of $h_n(\xi)=\sqrt{\frac{2}{L}}e^{-\frac{\beta}{2\eta}\xi}
\sin\left(\frac{n\pi\xi}{L}\right)$ and integration by parts yields
\begin{equation}
\int_0^\delta h_n(\xi)\,d\xi
={}\sqrt{\frac{2}{L}}\frac{L}{n\pi}
\left[
1-e^{-\frac{\beta\delta}{2\eta}}
\cos\left(\frac{n\pi\delta}{L}\right)
\right]
-\sqrt{\frac{2}{L}}\frac{\beta L}{2\eta n\pi}
\int_0^\delta e^{-\frac{\beta}{2\eta}\xi}
\cos\left(\frac{n\pi\xi}{L}\right)\,d\xi.
\end{equation}
Consequently, since $0<\delta\le L$, there exists a constant
$C>0$, independent of $n$, such that
\begin{equation}
\left|\int_0^\delta h_n(\xi)\,d\xi\right|
\le \frac{C}{n}.
\end{equation}
Moreover, recalling that
$\lambda_n=\alpha-\frac{\beta^2}{4\eta}
-\eta\left(\frac{n\pi}{L}\right)^2$,
we also have $|\lambda_n|\le Cn^2$. Therefore, we obtain
\begin{align}
\sum_{n=1}^{\infty}
\left|
\left(\int_0^\delta h_n(\xi)\,d\xi\right)
\langle x,A^*h_n^*\rangle_H
\right|
&\leq
C\sum_{n=1}^{\infty}
n|\langle x,h_n^*\rangle_H| \nonumber\\
&=
C\sum_{n=1}^{\infty}
\left(n^{2\theta}|\langle x,h_n^*\rangle_H|\right)n^{1-2\theta} \nonumber\\
&\leq
C
\left(\sum_{n=1}^{\infty}
n^{4\theta}|\langle x,h_n^*\rangle_H|^2\right)^{1/2}
\left(\sum_{n=1}^{\infty}n^{2-4\theta}\right)^{1/2}.
\end{align}

Since $\theta>\frac34$, we have $2-4\theta<-1$,
and therefore $\sum_{n=1}^{\infty}
n^{2-4\theta}<\infty$. It remains to estimate $\sum_{n=1}^{\infty}
n^{4\theta}|\langle x,h_n^*\rangle_H|^2$. Furthermore, since $x\in D_A(\theta,2)$, we have $\int_0^\infty
t^{1-2\theta}
\|AS(t)x\|_H^2\,dt
<\infty$.
By \eqref{rzbound},
\begin{equation}
\|AS(t)x\|_H^2
\geq
C
\sum_{n=1}^{\infty}
|\langle AS(t)x,h_n^*\rangle_H|^2
\nonumber\\
=
c
\sum_{n=1}^{\infty}
|\lambda_n|^2
e^{2\lambda_n t}
|\langle x,h_n^*\rangle_H|^2.
\end{equation}
Hence,
\begin{align}
\int_0^\infty
t^{1-2\theta}
\|AS(t)x\|_H^2\,dt
&\geq
c
\sum_{n=1}^{\infty}
|\lambda_n|^2
|\langle x,h_n^*\rangle_H|^2
\int_0^\infty
t^{1-2\theta}
e^{2\lambda_n t}\,dt.
\end{align}

Since $\lambda_n<0$, the change of variable $s=2|\lambda_n|t$ gives
\begin{equation}
\int_0^\infty
t^{1-2\theta}
e^{2\lambda_n t}\,dt
=
\frac{\Gamma(2-2\theta)}
{(2|\lambda_n|)^{2-2\theta}}.
\end{equation}
Therefore,
\begin{equation}
\int_0^\infty
t^{1-2\theta}
\|AS(t)x\|_H^2\,dt
\geq
c
\sum_{n=1}^{\infty}
|\lambda_n|^{2\theta}
|\langle x,h_n^*\rangle_H|^2.
\end{equation}
Since
\begin{equation}
d_1n^2
\leq
|\lambda_n|
\leq
d_2n^2,
\qquad n\geq1,
\end{equation}
for some constants $d_1,d_2>0$, it follows that
\begin{equation}
\sum_{n=1}^{\infty}
n^{4\theta}
|\langle x,h_n^*\rangle_H|^2
\leq
C\|x\|_{D_A(\theta,2)}^2.
\end{equation}
Combining the above estimates yields
\begin{equation}
\sum_{n=1}^{\infty}
\left|
\left(
\int_0^\delta h_n(\xi)\,d\xi
\right)
\langle x,A^*h_n^*\rangle_H
\right|
\leq
C\|x\|_{D_A(\theta,2)}
<\infty.
\end{equation}
The linearity of $\Lambda$ is immediate, and the proof is complete.
\end{proof}

We now consider the solution to the SPDE \eqref{eq:spde_positive_side} and establish conditions under which it takes values in $D_A(\theta,2), \theta>\frac{3}{4}$. We impose the following assumptions on the drift and diffusion coefficients.
\begin{assumption}
\label{ass:regularity}
Let $\frac14<\alpha<\frac12$. The following conditions hold:
\begin{enumerate}[(i)]
\item The initial condition satisfies $u_0\in D_A\left(\alpha+\frac12,2\right)$.
\item The mapping $V:[0,T]\times\Omega\times H
\longrightarrow
D_A\left(\alpha+\frac12,2\right)$
is $\mathcal{P}\otimes\mathcal{B}(H)$-measurable, where $\mathcal F_\mathcal{P}$ denotes the progressively measurable $\sigma$-field on $[0,T]\times\Omega$ associated with $(\mathcal F_t)_{t\in[0,T]}$. Moreover, there exists a constant $L_V>0$ such that 
\begin{equation*}
\|V(t,\omega,x)-V(t,\omega,y)\|_{D_A\left(\alpha+\frac12,2\right)}
\leq
L_V\|x-y\|_H
\end{equation*}
for all $x,y\in H$, $t\in[0,T]$, and almost every $\omega\in\Omega$. In addition,
\begin{equation*}
\mathbb{E}\int_0^T
\|V(t,\omega,0)\|_{D_A\left(\alpha+\frac12,2\right)}^2\,dt
<\infty.
\end{equation*}

\item The mapping
$B\in\mathcal L\left(H,L_2(H_0;D_A(\alpha,2))\right)$
satisfies 
\begin{equation*}
\|B(x)\|_{L_2(H_0;D_A(\alpha,2))}^2
\leq \varrho\|x\|_H^2
\leq \rho\|x\|_{D_A\left(\alpha+\frac12,2\right)}^2,
\end{equation*}
where $\varrho,\rho>0$ are constants satisfying
$\rho<1-2\alpha$.
\end{enumerate}
\end{assumption}

\begin{theorem}
\label{prop:regularity}
Suppose that Assumption~\ref{ass:regularity} holds. Then the SPDE
\eqref{eq:spde_positive_side} admits a unique mild solution 

$u\in
\mathcal {N}_{\mathcal{F}}\left(
0,T;
D_A\left(\alpha+\frac12,2\right)
\right)$.

\end{theorem}
\begin{proof}
The proof is a slight modification of the contraction argument in
Theorem~6.21 of \cite{da2014stochastic}. Define
\begin{equation}
\Phi(u)(t)
=
S(t)u_0
+
\int_0^t
S(t-s)V(s,\omega,u_s)\,ds
+
\int_0^t
S(t-s)B(u_s)\,dW_s.
\end{equation}
By the estimates used in Theorem~6.21 of \cite{da2014stochastic} and
Assumption~\ref{ass:regularity},
\begin{equation}
\|\Phi(u)-\Phi(v)\|
\leq
\left(
\sqrt{\frac{\rho}{1-2\alpha}}
+
CT
\right)
\|u-v\|.
\end{equation}
where $C>0$ is independent of $T$. Since $\frac{\rho}{1-2\alpha}<1$, $\Phi$ is a contraction when $T_0>0$ is sufficiently small. Hence, a
unique mild solution exists on $[0,T_0]$. Repeating the same argument
on successive intervals of length $T_0$ yields the unique solution on
$[0,T]$.
\end{proof}
\begin{remark}
Consistent with the stylized facts of the limit order book, we require
$u_t^a\geq 0$ (and $u_t^b\leq 0$). Here, the order $\geq$ is the standard
pointwise order on the corresponding function spaces.
Such positivity requirements usually require additional and independent
assumptions on the coefficients of the SPDE
\cite{Milian2002,GoncharukKotelenez1998}.
For example, based on \cite{GoncharukKotelenez1998}, in addition to
\Cref{ass:regularity}, suppose that $u_0\geq 0$, $V$ is bounded and
satisfies $V(t,\omega,0)\geq 0$, and the noise coefficient is given by the
pointwise multiplication operator $(Bf)(x)=\sigma(x)f(x), \forall f \in H$, where $\sigma$
is a bounded function. Then the positivity result in
\cite{GoncharukKotelenez1998} applies and yields $u_t\geq 0$ for all
$t\in[0,T]$. The analysis in this paper does not directly rely on the
specific conditions ensuring positivity. We therefore do not impose
explicitly a separate set of such conditions, but assume throughout that
the order density processes satisfy the required positivity properties.
\end{remark}
The next proposition shows that, under the present framework, the market depth process $D$ is an It\^o process.
\begin{proposition}\label{prop:depth_dynamics}
Suppose that \Cref{ass:regularity} holds and let $u$ be the
mild solution obtained in \Cref{prop:regularity}. Then the
market depth process $D$ satisfies
\begin{align}
D_t
=
D_0
+
\int_0^t
\left[
\Lambda(u_s)
+
\left\langle
V(s,\omega,u_s),\mathbf{1}_{[0,\delta]}
\right\rangle_H
\right]ds +
\int_0^t
B(u_s)^*\mathbf{1}_{[0,\delta]}
\,dW_s,
\end{align}
where $B(u_s)^*\in L_2(H,H_0)$ denotes the adjoint of $B(u_s)\in L_2(H_0,H)$.
\end{proposition}

\begin{proof}
Let
\begin{equation}
\delta_n:=\int_0^\delta h_n(\xi)\,d\xi.
\end{equation}
For each $n\geq1$, since $h_n^*\in D(A^*)$, the weak formulation gives
\begin{align}
\langle u_t,h_n^*\rangle_H
=
\langle u_0,h_n^*\rangle_H
+
\int_0^t
\langle u_s,A^*h_n^*\rangle_H\,ds
+
\int_0^t
\langle V(s,\omega,u_s),h_n^*\rangle_H\,ds
+
\int_0^t
B(u_s)^*h_n^*\,dW_s.
\end{align}
Multiplying by $\delta_n$ and summing over $n=1,\ldots,N$ yields 
\begin{align}
\sum_{n=1}^N \delta_n\langle u_t,h_n^*\rangle_H
&=
\sum_{n=1}^N \delta_n\langle u_0,h_n^*\rangle_H
+
\int_0^t \sum_{n=1}^N
\delta_n\langle u_s,A^*h_n^*\rangle_H\,ds \nonumber\\
&\quad+
\int_0^t \sum_{n=1}^N
\delta_n\langle V(s,\omega,u_s),h_n^*\rangle_H\,ds
+
\int_0^t B(u_s)^*
\left(\sum_{n=1}^N\delta_n h_n^*\right)dW_s.
\end{align}

For each fixed $t\in[0,T]$, the following convergences hold in
$L^2(\Omega)$:
\begin{align}
\sum_{n=1}^N
\delta_n\langle u_t,h_n^*\rangle_H
&\longrightarrow D_t,
\\
\sum_{n=1}^N
\delta_n\langle u_0,h_n^*\rangle_H
&\longrightarrow D_0,
\\
\int_0^t
\sum_{n=1}^N
\delta_n\langle u_s,A^*h_n^*\rangle_H\,ds
&\longrightarrow
\int_0^t\Lambda(u_s)\,ds,
\\
\int_0^t
\sum_{n=1}^N
\delta_n\langle V(s,\omega,u_s),h_n^*\rangle_H\,ds
&\longrightarrow
\int_0^t
\left\langle
V(s,\omega,u_s),\mathbf{1}_{[0,\delta]}
\right\rangle_H\,ds,
\\
\int_0^t
B(u_s)^*
\left(
\sum_{n=1}^N\delta_n h_n^*
\right)
\,dW_s
&\longrightarrow
\int_0^t
B(u_s)^*\mathbf{1}_{[0,\delta]}
\,dW_s.
\end{align}
Here we use \Cref{prop:regularity}, the Riesz basis expansion, the
dominated convergence theorem, and the It\^o isometry.
\end{proof}
Applying \Cref{prop:depth_dynamics} to the ask and bid sides, respectively, the market depth processes satisfy 
\begin{align}
dD_t^a
&=
\left[
\Lambda_a(u_t^a)
+
\left\langle
V^a(t,\omega,u_t^a),
\mathbf{1}_{[0,\delta]}
\right\rangle_{H^a}
\right]dt
+
B^a(u_t^a)^*\mathbf{1}_{[0,\delta]}\,dW_t^a,
\\
dD_t^b
&=
-\left[
\Lambda_b(u_t^b)
+
\left\langle
V^b(t,\omega,u_t^b),
\mathbf{1}_{[-\delta,0]}
\right\rangle_{H^b}
\right]dt
-
B^b(u_t^b)^*\mathbf{1}_{[-\delta,0]}\,dW_t^b.
\end{align}
Therefore, the above representation of the market depth dynamics yields a tractable form of the mid-price dynamics. In view of \eqref{eq:mid_price}, we obtain 
\begin{align}
dF_t
&=
-\kappa
\left[
\frac{
\Lambda_b(u_t^b)
+
\left\langle
V^b(t,\omega,u_t^b),
\mathbf{1}_{[-\delta,0]}
\right\rangle_{H^b}
}{D_t^b}
+
\frac{
\Lambda_a(u_t^a)
+
\left\langle
V^a(t,\omega,u_t^a),
\mathbf{1}_{[0,\delta]}
\right\rangle_{H^a}
}{D_t^a}
\right]dt
\nonumber\\
&\quad
-\kappa
\frac{
B^b(u_t^b)^*\mathbf{1}_{[-\delta,0]}
}{D_t^b}
\,dW_t^b
-\kappa
\frac{
B^a(u_t^a)^*\mathbf{1}_{[0,\delta]}
}{D_t^a}
\,dW_t^a.\label{midprice-SDE}
\end{align}
\begin{remark}
An alternative, approximate method to address the case where
\Cref{eq:general_lob} does not satisfy the regularity condition
\Cref{ass:regularity}, so that its solution does not belong to the
required interpolation space, is as follows. We consider only the ask side, as the bid side can be treated analogously. Since $D(A_a^*)$ is a dense subspace of $H^a$, we can choose
$\psi^a\in D(A_a^*)$ approximating $\mathbf{1}_{[0,\delta]}$ and use
$\langle u_t^a,\psi^a\rangle_{H^a}$ to approximate
$\langle u_t^a,\mathbf{1}_{[0,\delta]}\rangle_{H^a}$. By the
definition of a mild (weak) solution, $\langle u_t^a,\psi^a\rangle_{H^a}$
then satisfies the SDE
\begin{equation}
d\langle u_t^a,\psi^a\rangle_{H^a}
=
\left[
\langle u_t^a,A_a^*\psi^a\rangle_{H^a}
+
\langle V^a(t,\omega,u_t^a),\psi^a\rangle_{H^a}
\right]dt
+
\langle B^a(u_t^a)\,dW_t^a,\psi^a\rangle_{H^a}.
\end{equation}
\end{remark}

\section{$N$-Player Game of Optimal Trading}\label{NPgm}

We now formulate an $N$-player game of optimal trading associated with the limit order book
model introduced in the previous section. Following a trading model similar
to that in \cite{JaimungalBook2015}, each trader chooses a trading rate and
aims to minimize the cost of executing her position over the trading horizon
$[0,T]$. In addition, the traders interact through the market, since their aggregate
trading activity, modeled by the average trading rate $v^{(N)}$, 
affects the limit order book densities and the mid-price process $F^{(N)}$.

\subsection{Optimal Trading Problem} \label{gese}
We consider a population of agents who executing market orders to
buy or sell the asset. For each trader $i=1,\ldots,N$, let
$q^i=(q_t^i)_{t\in[0,T]}$ denote her inventory process and let
$v^i=(v_t^i)_{t\in[0,T]}$ denote her trading rate. We use the convention that $v_t^i>0$ corresponds to buying and $v_t^i<0$
corresponds to selling. The inventory dynamics are
given by 
\begin{equation}
dq_t^i
=
v_t^i\,dt
+
\sigma^q\,d\beta_t^i, \quad \sigma^q>0,
\quad q^i_0\in \mathbb{R}_{>0},\quad
i=1,\ldots,N,
\end{equation}
where $\beta^i$, $i=1,\ldots,N$, are independent one-dimensional Brownian
motions, and are also independent of $W^a$ and $W^b$. As argued in \cite{carmona2023optimal}, inventory is not differentiable and has a Brownian component, which motivates the noise in our inventory dynamics and also provides mathematical generality. Without this idiosyncratic noise, the entire analysis still goes through.
\begin{remark}
To construct all the driving noises on the same probability space, we can follow
the construction in \cite{liu2025hilbert} and partition a sequence of mutually independent
one-dimensional Brownian motions into three disjoint subsequences. The first
two subsequences are used to construct the $Q_a$- and $Q_b$-Wiener processes
$W^a$ and $W^b$, respectively, while the third is denoted by
$(\beta^i)_{i\geq1}$. Consequently, $W^a$, $W^b$, and
$(\beta^i)_{i\geq1}$ are mutually independent.
\end{remark}

For each $N\geq1$, we define the filtration
$\mathcal F^{[N]}=(\mathcal F_t^{[N]})_{t\in[0,T]}$ by
\begin{equation}
\mathcal F_t^{[N]}
:=
\sigma\left(
W_s^a,W_s^b,\beta_s^1,\ldots,\beta_s^N:
0\leq s\leq t
\right)\vee\mathcal N,
\qquad t\in[0,T],
\end{equation}
where $\mathcal N$ denotes the collection of $\mathbb P$-null sets.
Thus,
$(\Omega,\mathcal F,(\mathcal F_t^{[N]})_{t\in[0,T]},\mathbb P)$
is the filtered probability space associated with the $N$-player game. The admissible controls are taken from the space $\mathcal M_{\mathcal F^{[N]}}(0,T)$. 

The process $F^{N}$ represents the market mid-price, which is determined
by the ask- and bid-side limit order book densities as introduced in
Section~\ref{lobd}, in turn, affected by the trading activities $v^1,\ldots,v^N$ of all market
participants. Therefore, the traders interact through their collective
impact on the limit order book and, consequently, on the mid-price. A
specific form of this dependence will be introduced in \Cref{sec:maro-micro}.

The trading activity of each trader also has a temporary impact on her
execution price. The execution price of trader $i$ is given by
\begin{equation}
S_t^i
=
F_t^{N}
+
a v_t^i,\quad S^i_0 \in \mathbb{R}_{>0},
\quad
i=1,\ldots,N,
\label{ExecutionPrice}
\end{equation}
where $a\geq0$ measures the strength of the temporary price impact.

Let $Z^i=(Z_t^i)_{t\in[0,T]}$ denote the cash process of trader $i$.
Its dynamics are given by 
\begin{equation}
dZ_t^i
=
-S_t^i\,dq_t^i,\quad Z^i_0 \in \mathbb{R}_{>0},
\quad
i=1,\ldots,N.
\label{minorCashProcess}
\end{equation}
Thus, a purchase decreases the trader's cash position, while a sale
increases it.

Each trader aims to trade by posting market orders while controlling inventory risk. We consider the cost functional
\begin{equation}
J_i(v^i,v^{-i})
=
\mathbb{E}
\left[
-Z_T^i
-(q_T^i-r)
\left(
F_T^{N}-\xi(q_T^i-r)
\right)
+
\int_0^T
\phi(q_s^i-r)^2\,ds
\right],
\qquad
i=1,\ldots,N,
\label{minorLiqCostFunction}
\end{equation}
where $\xi>0$, $\phi>0$ are constants, and $r$ satisfies $0 \leq r \leq q^i_0$ and denotes the target inventory level for trader $i$. Moreover, $v^{-i}
:=
(v^1,\ldots,v^{i-1},v^{i+1},\ldots,v^N)
$ denotes the trading strategies of all traders other than trader $i$.

The $N$-player game consists of $N$ traders, where each trader $i$ chooses a strategy $v^i$ to minimize
$J_i(v^i,v^{-i})$, given the strategies of the other traders $v^{-i}$.
\subsection{Limit Order Book Dynamics with Strategic Traders and Price Formation}\label{sec:maro-micro}

We now revisit the general framework introduced in \Cref{lobd} and specify how it can be formulated in a game-theoretic setting for optimal trading problems. The limit order book model introduced there provides a description of market liquidity through the ask- and bid-side density
processes. We now consider an $N$-player game
describing the strategic behavior of market participants submitting market orders. Since the limit order
book density is affected by the combined buying and selling activity of
market participants, it is natural to describe the impact of the traders
through their aggregate buy and sell market orders. We combine the two models and represent the limit order book dynamics by the 
system of stochastic evolution equations given as
\begin{align}
du_t^{a,N}
&=
\left[
A_au_t^{a,N}
+
G^a\left(
\sum_{i=1}^{N_a} v_t^{i,+},
\sum_{i=1}^{N_b} v_t^{i,-},
u_t^a,t
\right)
+J^a(u_t^{a,N},t)\right]dt
+
B^a(u_t^{a,N})\,dW_t^a,
\\
du_t^{b,N}
&=
\left[
A_b u_t^{b,N}
+
G^b\left(
\sum_{i=1}^{N_a} v_t^{i,+},
\sum_{i=1}^{N_b} v_t^{i,-},
u_t^{b,N},t
\right)+J^b(u_t^{b,N},t)
\right]dt
+
B^b(u_t^{b,N})\,dW_t^b.
\\
&u_0^{a,N}  \in D_{A_a}\left(\alpha+\frac12,2\right),
\qquad u_0^{a,N}(x)\geq 0,\quad \forall x \in [0,L],
\\
&u_0^{b,N}  \in D_{A_b}\left(\alpha+\frac12,2\right),
\qquad u_0^{b,N}(x)\leq 0,\quad \forall x \in [-L,0].
\end{align}
Here, $N_a$ and $N_b$ denote the numbers of traders submitting buy and sell market orders, respectively, with $N = N_a + N_b$. The quantities $\sum_{i=1}^{N_a} v_t^{i,+}$ and $\sum_{i=1}^{N_b} v_t^{i,-}$ represent the corresponding aggregate buy and sell market orders at time $t$. The terms $G^a$ and $G^b$ model the effects of these $N$ traders’ market orders on the ask- and bid-side densities, respectively. These traders are also assumed to post, replace, and cancel limit orders, which are taken to be known and are not modeled as part of an optimization problem in this work. The terms $J^a$ and $J^b$ model additional inflows and outflows of orders on the ask and bid sides, respectively. For example, they may represent the aggregate order flow generated by the same traders posting new limit orders, or by participants outside the population of agents considered in the game.

In
the present setting, buy and sell orders do not primarily interact with
each other through their empirical averages; rather, their cumulative
effect is transmitted to the opposite side of the limit order book. Consequently, the limiting equations as the number of traders $N$ tends to infinity may not be well defined, which prevents a direct application of the standard mean field game (MFG) methodology. Moreover, directly replacing the aggregate order flow by its population average
\begin{equation}
\frac{1}{N}\sum_{i=1}^{N} v_t^i
=
\frac{1}{N}
\left(
\sum_{i=1}^{N_a} v_t^{i,+}
+
\sum_{i=1}^{N_b} v_t^{i,-}
\right),
\qquad
N=N_a+N_b.
\end{equation}
may substantially underestimate its impact on the limit order book density, particularly when the number of market participants is large. For example, suppose that there are $N=100$ market participants, with
$60$ submitting buy market orders and $40$ submitting sell market orders,
each of unit size. The resulting market order imbalance is then $20$,
whereas its population average is only $\frac{60-40}{100}=0.2$. Thus, although the aggregate imbalance represents the effect of $20$ unit
orders on the order book density, direct averaging reduces this effect to
$0.2$, which is even smaller than the size of a single order.

Motivated by this observation, we proceed heuristically to derive a system that is amenable to the MFG methodology while retaining a reasonable representation of the underlying market dynamics. We assume that $G^a$ and $G^b$ take the form
\begin{align} \label{GaGb}
G^a\left(
\sum_{i=1}^{N_a}v_t^{i,+}
+
\sum_{i=1}^{N_b}v_t^{i,-},
u_t^a,t
\right),
\qquad
G^b\left(
\sum_{i=1}^{N_a}v_t^{i,+}
+
\sum_{i=1}^{N_b}v_t^{i,-},
u_t^b,t
\right).
\end{align}
This means that the aggregate effect of the $N$ traders' buy and sell market orders on the limit order book density is determined solely through their market order imbalance. Also, note that $\sum_{i=1}^{N_a}v_t^{i,+}
+
\sum_{i=1}^{N_b}v_t^{i,-}=\sum_{i=1}^{N} v_t^i$.
Thus, the ask- and bid-side densities take the form 
\begin{align} \label{dylobN}
u_t^{a,N}
&=
S_a(t)Nu^a_{0}
+
\int_0^t
S_a(t-s)
\left[
G^a\left(
\sum_{i=1}^{N} v_s^i,
u_s^{a,N},s
\right)
+
J^a(u_s^{a,N},s)
\right]ds
\nonumber\allowdisplaybreaks\\
&\quad+
\int_0^t
S_a(t-s)B^a(u_s^{a,N})\,dW_s^a,
\qquad u^a\in H,
\allowdisplaybreaks\\
u_t^{b,N}
&=
S_b(t)Nu_0^b
+
\int_0^t
S_b(t-s)
\left[
G^b\left(
\sum_{i=1}^{N} v_s^i,
u_s^{b,N},s
\right)
+
J^b(u_s^{b,N},s)
\right]ds
\nonumber\allowdisplaybreaks\\
&\quad+
\int_0^t
S_b(t-s)B^b(u_s^{b,N})\,dW_s^b,
\qquad u^b\in H,
\end{align}
where the initial conditions $u_0^{a,N}=Nu^a_0$ and $u_0^{b,N}=Nu^b_0$, with $u_0^a\in H^a$, $u_0^b\in H^b$ being fixed initial densities that do not depend on $N$, reflect the fact that the volume of limit orders at each price level increases with the number of market participants. From \eqref{midprice-SDE}, the mid-price dynamics in this setting are given by
\begin{align} \label{F^N}
F_t^N
&=
F_0
-\kappa
\int_0^t
\left[
\frac{
\Lambda_b(u_s^{b,N})
+
\left\langle
G^b\left(
\sum_{i=1}^{N}v_s^i,
u_s^{b,N},s
\right)
+
J^b(u_s^{b,N},s),
\mathbf{1}_{[-\delta,0]}
\right\rangle_{H^b}
}{
-\left\langle
u_s^{b,N},
\mathbf{1}_{[-\delta,0]}
\right\rangle_{H^b}
}
\right.
\nonumber\\
&\qquad\left.
+
\frac{
\Lambda_a(u_s^{a,N})
+
\left\langle
G^a\left(
\sum_{i=1}^{N}v_s^i,
u_s^{a,N},s
\right)
+
J^a(u_s^{a,N},s),
\mathbf{1}_{[0,\delta]}
\right\rangle_{H^a}
}{
\left\langle
u_s^{a,N},
\mathbf{1}_{[0,\delta]}
\right\rangle_{H^a}
}
\right]ds
\nonumber\\
&\quad
-\kappa
\int_0^t
\frac{
B^b(u_s^{b,N})^*
\mathbf{1}_{[-\delta,0]}
}{
-\left\langle
u_s^{b,N},
\mathbf{1}_{[-\delta,0]}
\right\rangle_{H^b}
}
\,dW_s^b-\kappa
\int_0^t
\frac{
B^a(u_s^{a,N})^*
\mathbf{1}_{[0,\delta]}
}{
\left\langle
u_s^{a,N},
\mathbf{1}_{[0,\delta]}
\right\rangle_{H^a}
}
\,dW_s^a.
\end{align}
Assume further that $G^a$, $G^b$, $J^a$, and $J^b$ are homogeneous of degree one, so that, for any $\lambda>0$,
\begin{align} \label{homoge}
G^a(\lambda v,\lambda u,t)
&=
\lambda G^a(v,u,t),
&
G^b(\lambda v,\lambda u,t)
&=
\lambda G^b(v,u,t),
\\
J^a(\lambda u,t)
&=
\lambda J^a(u,t),
&
J^b(\lambda u,t)
&=
\lambda J^b(u,t).
\end{align}
Moreover, $B_l(\cdot)$, $l\in\{a,b\}$, is a linear operator and hence is also homogeneous of degree one. Dividing the ask- and bid-side densities by the number of traders $N$ and defining
\begin{equation} \label{res}
u_t^{a,(N)}
:=
\frac{u_t^{a,N}}{N},
\qquad
u_t^{b,(N)}
:=
\frac{u_t^{b,N}}{N},
\qquad
v_t^{(N)}
:=
\frac{1}{N}\sum_{i=1}^N v_t^i,
\end{equation}
we obtain
\begin{align}
u_t^{a,(N)}
&=
S_a(t)u^a_{0}
+
\int_0^t
S_a(t-s)
\left[
G^a\left(
v_s^{(N)},
u_s^{a,(N)},
s
\right)
+
J^a\left(
u_s^{a,(N)},s
\right)
\right]ds
\nonumber\\
&\quad+
\int_0^t
S_a(t-s)B^a\left(u_s^{a,(N)}\right)\,dW_s^a,
 \label{avdylobN-1}\\
u_t^{b,(N)}
&=
S_b(t)u^b_{0}
+
\int_0^t
S_b(t-s)
\left[
G^b\left(
v_s^{(N)},
u_s^{b,(N)},
s
\right)
+
J^b\left(
u_s^{b,(N)},s
\right)
\right]ds
\nonumber\\
&\quad+
\int_0^t
S_b(t-s)B^b\left(u_s^{b,(N)}\right)\,dW_s^b,\label{avdylobN-2}
\end{align}
where the processes $u_t^{a,(N)}$ and $u_t^{b,(N)}$ are the ask- and bid-side order densities normalized by the number of traders. They represent the liquidity available per market participant, and, unlike $u_t^{a,N}$ and $u_t^{b,N}$, they remain of order one as $N\to\infty$ and have a well-defined limit. We therefore refer to \eqref{avdylobN-1}--\eqref{avdylobN-2} as the \emph{normalized} (or \emph{per-trader}) densities.
The corresponding mid-price process associated with the 
normalized density system, denoted by $F^{(N)}$, satisfies the dynamics 
\begin{align}
F_t^{(N)} \label{avF^N}
&=
F_0
-\kappa
\int_0^t
\left[
\frac{
\Lambda_b(u_s^{b,(N)})
+
\left\langle
G^b\left(
v_s^{(N)},
u_s^{b,(N)},s
\right)
+
J^b(u_s^{b,(N)},s),
\mathbf{1}_{[-\delta,0]}
\right\rangle_{H^b}
}{
-\left\langle
u_s^{b,(N)},
\mathbf{1}_{[-\delta,0]}
\right\rangle_{H^b}
}
\right.
\nonumber\allowdisplaybreaks\\
&\qquad\left.
+
\frac{
\Lambda_a(u_s^{a,(N)})
+
\left\langle
G^a\left(
v_s^{(N)},
u_s^{a,(N)},s
\right)
+
J^a(u_s^{a,(N)},s),
\mathbf{1}_{[0,\delta]}
\right\rangle_{H^a}
}{
\left\langle
u_s^{a,(N)},
\mathbf{1}_{[0,\delta]}
\right\rangle_{H^a}
}
\right]ds
\nonumber\allowdisplaybreaks\\
&\quad
-\kappa
\int_0^t
\frac{
B^b(u_s^{b,(N)})^*
\mathbf{1}_{[-\delta,0]}
}{
-\left\langle
u_s^{b,(N)},
\mathbf{1}_{[-\delta,0]}
\right\rangle_{H^b}
}
\,dW_s^b-\kappa
\int_0^t
\frac{
B^a(u_s^{a,(N)})^*
\mathbf{1}_{[0,\delta]}
}{
\left\langle
u_s^{a,(N)},
\mathbf{1}_{[0,\delta]}
\right\rangle_{H^a}
}
\,dW_s^a.
\end{align}
Moreover, by \eqref{homoge}--\eqref{res} and the linearity of the remaining
terms, including $B_l(\cdot)$ and $\Lambda_l(\cdot)$ for $l\in\{a,b\}$, the numerator and the denominator of each term in the dynamics of $F^{N}$ scale by the same factor $N$ relative to those of $F^{(N)}$. Hence, this factor cancels, and we obtain 
\begin{equation}
F_t^N=F_t^{(N)}, \quad \mathbb{P}-\text{a.s.},
\qquad t\in[0,T].
\end{equation}
Thus, \eqref{avdylobN-1}--\eqref{avF^N} constitute the normalized (per-trader) system associated with
\eqref{dylobN}--\eqref{F^N}, which represents the real market system.
The two systems generate the same mid-price process, while the normalized
system is driven by the average trading rate $v_t^{(N)}=\frac{1}{N}\sum_{i=1}^N v_t^i$ instead of the aggregate (total) trading rate $\sum_{i=1}^N v_t^i$, for $t\in[0,T]$. 

The following example illustrates how the mappings $G^a$ and $G^b$ in \eqref{GaGb} may be constructed.
\begin{example}
Suppose that the effects of $G^a$ and $G^b$ are described by mappings
$f^a:\mathbb{R}\times H^a\to H^a$ and
$f^b:\mathbb{R}\times H^b\to H^b$,
which are linear in their first arguments. For instance, on the ask side, the buy market orders remove the amount $f^a \left(
\sum_{i=1}^{N_a}v_t^{i,+},u_t^a \right)$
from the ask-side density. At the same time, the amount $f^a \left(
\min\left\{
\sum_{i=1}^{N_a}v_t^{i,+},
-\sum_{i=1}^{N_b}v_t^{i,-}
\right\},
u_t^a
\right)$
is replenished, representing the part of the buy and sell market orders that offsets each other. Therefore, the part of the change in the ask-side density resulting from the market order executions of the $N$ traders is given by
\begin{align} \label{faex}
&G^a:=f^a\left(
\min\left\{
\sum_{i=1}^{N_a}v_t^{i,+},
-\sum_{i=1}^{N_b}v_t^{i,-}
\right\},
u_t^a
\right)
-
f^a\left(
\sum_{i=1}^{N_a}v_t^{i,+},
u_t^a
\right)
\nonumber\\
&\qquad=
-f^a\left(
\left(
\sum_{i=1}^{N_a}v_t^{i,+}
+
\sum_{i=1}^{N_b}v_t^{i,-}
\right)^+,
u_t^a
\right)
=
-f^a\left(
\left(\sum_{i=1}^{N}v_t^i\right)^+,
u_t^a
\right).
\end{align}
Similarly, for the bid side, we have 
\begin{align} \label{fbex}
&G^b:=f^b\left(
\min\left\{
\sum_{i=1}^{N_a}v_t^{i,+},
-\sum_{i=1}^{N_b}v_t^{i,-}
\right\},
u_t^b
\right)
-
f^b\left(
-\sum_{i=1}^{N_b}v_t^{i,-},
u_t^b
\right)
\nonumber\\
&\qquad=
-f^b\left(
\left(
\sum_{i=1}^{N_a}v_t^{i,+}
+
\sum_{i=1}^{N_b}v_t^{i,-}
\right)^-,
u_t^b
\right)
=
-f^b\left(
\left(\sum_{i=1}^{N}v_t^i\right)^-,
u_t^b
\right).
\end{align}
Here, $(x)^{+}:=\max\{x,0\}$ and $(x)^{-}:=\max\{-x,0\}$ for $x\in\mathbb{R}$, which correspond to the positive and negative parts of $x$, so that $x=x^{+}-x^{-}$ and $x^{\pm}\ge 0$. As illustrated by \eqref{faex} and \eqref{fbex}, the effects of market orders on the ask- and bid-side densities may initially depend on the aggregate buy and sell market orders separately. However, under the particular mechanism described above, this dependence reduces to the aggregate market order imbalance alone.
\end{example}

\subsection{Truncated Density-Price System}
\label{sec:truncated_density_price}
The system \eqref{avdylobN-1}--\eqref{avF^N} is associated with the following
general setting. Let $(\Omega,\mathcal F,(\mathcal F_t)_{t\in[0,T]},\mathbb P)$ be a filtered
probability space, and let
$W^a,W^b$ be a $\mathcal F$-adapted $Q$-Wiener process. For an
$\mathcal F_{P}$-measurable process $g$, define the mid-price process by
\begin{align}
\label{Ftor}
F_t
&=
F_0
-\kappa
\int_0^t
\left[
\frac{
\Lambda_b(u_s^b)
+
\left\langle
G^b\left(
g_s,u_s^b,s
\right)
+
J^b(u_s^b,s),
\mathbf{1}_{[-\delta,0]}
\right\rangle_{H^b}
}{
-\left\langle
u_s^b,
\mathbf{1}_{[-\delta,0]}
\right\rangle_{H^b}
}
\right.
\nonumber\\
&\qquad\left.
+
\frac{
\Lambda_a(u_s^a)
+
\left\langle
G^a\left(
g_s,u_s^a,s
\right)
+
J^a(u_s^a,s),
\mathbf{1}_{[0,\delta]}
\right\rangle_{H^a}
}{
\left\langle
u_s^a,
\mathbf{1}_{[0,\delta]}
\right\rangle_{H^a}
}
\right]ds
\nonumber\\
&\quad
-\kappa
\int_0^t
\frac{
B^b(u_s^b)^*\mathbf{1}_{[-\delta,0]}
}{
-\left\langle
u_s^b,
\mathbf{1}_{[-\delta,0]}
\right\rangle_{H^b}
}
\,dW_s^b
-\kappa
\int_0^t
\frac{
B^a(u_s^a)^*\mathbf{1}_{[0,\delta]}
}{
\left\langle
u_s^a,
\mathbf{1}_{[0,\delta]}
\right\rangle_{H^a}
}
\,dW_s^a,
\end{align}
where the limit order density processes $u^a$ and $u^b$ are given by
\begin{align}
u_t^a
&=
S_a(t)u^a_{0}
+
\int_0^t
S_a(t-s)
\left[
G^a\left(
g_s,u_s^a,s
\right)
+
J^a\left(
u_s^a,s
\right)
\right]ds
+
\int_0^t
S_a(t-s)B^a(u_s^a)\,dW_s^a,
\label{loblim-1}\\
u_t^b
&=
S_b(t)u^b_{0}
+
\int_0^t
S_b(t-s)
\left[
G^b\left(
g_s,u_s^b,s
\right)
+
J^b\left(
u_s^b,s
\right)
\right]ds
+
\int_0^t
S_b(t-s)B^b(u_s^b)\,dW_s^b.\label{loblim-2}
\end{align}
The system \eqref{Ftor}--\eqref{loblim-2} consists of two semilinear SPDEs
for the limit order density processes, coupled with a highly nonlinear mid-price
equation involving singular terms. Consequently, obtaining an analytical
solution to the coupled system is generally intractable, particularly within
the MFG framework, where the input process $g$ is itself determined through
an equilibrium consistency condition. Thus, we introduce a truncated version of the mid-price process given by
\begin{align}\label{trf}
F_t
&=
F_0
-
\kappa
\int_0^t
\left[
\frac{
T_{-C_U,C_U}\Lambda_b(u_s^b)
+
\langle \mathbf{1}_{[-\delta,0]},J_s^b\rangle
+
\langle \mathbf{1}_{[-\delta,0]},G^b(g_s,u_s^b)\rangle
}{
T_{C_L,C_U}\left(
-\langle \mathbf{1}_{[-\delta,0]},u_s^b\rangle
\right)
}
\right.
\nonumber\\
&\qquad\left.
+
\frac{
T_{-C_U,C_U}\Lambda_a(u_s^a)
+
\langle \mathbf{1}_{[0,\delta]},J_s^a\rangle
+
\langle \mathbf{1}_{[0,\delta]},G^a(g_s,u_s^a)\rangle
}{
T_{C_L,C_U}\langle \mathbf{1}_{[0,\delta]},u_s^a\rangle
}
\right]ds
\nonumber\\
&\qquad
-
\kappa
\int_0^t
\Pi_R^b\!\left(
\frac{
B^b(u_s^b)^*\mathbf{1}_{[-\delta,0]}
}{
T_{C_L,C_U}\left(
-\langle \mathbf{1}_{[-\delta,0]},u_s^b\rangle
\right)
}
\right)dW_s^b
-
\kappa
\int_0^t
\Pi_R^a\!\left(
\frac{
B^a(u_s^a)^*\mathbf{1}_{[0,\delta]}
}{
T_{C_L,C_U}\langle \mathbf{1}_{[0,\delta]},u_s^a\rangle
}
\right)dW_s^a,
\end{align}
where $C_U>C_L>0$ are fixed truncation bounds, and, for any
$l<u$, the truncation function $T_{l,u}:\mathbb R\to\mathbb R$
is defined by
\begin{equation}
T_{l,u}(x):=
\begin{cases}
l, & x<l,\\
x, & l\le x\le u,\\
u, & x>u,
\end{cases}
\qquad x\in\mathbb R.
\end{equation}
Moreover, $\Pi_R^a$ and $\Pi_R^b$ denote the metric projections
onto the closed balls of radius $R$ in the corresponding Hilbert spaces.

The order density processes $u^a$ and $u^b$ remain unchanged; only the
expression for the mid-price process $F$ is modified by the truncation. Since all the coefficients are bounded under \eqref{trf}, whenever the
truncated mid-price process $F$ is well defined, its second moment admits the
uniform bound
\begin{equation} \label{boundF}
\sup_{t\in[0,T]}\mathbb E\left[|F_t|^2\right]
\leq C_F,
\end{equation}
where $C_F$ is independent of the underlying filtered probability space, the
choice of filtration, and the input process $g$. In particular, \eqref{boundF} remains valid when $g$ is adapted to a filtration larger than $\mathcal{F}$.

The truncation is introduced for technical reasons. First, it ensures the
analytical solvability of the mean field game on a small time interval.
Moreover, the uniform bound in \eqref{boundF} is needed to establish the
$\epsilon$-Nash equilibrium result. Since the truncated version of the
mid-price dynamics can be made arbitrarily close to the untruncated version
by choosing sufficiently large and sufficiently small truncation bounds,
respectively, we use the same notation for both versions.

 From a practical point of view, the truncation regularizes the response of the mid-price to the state of the limit order book. While the truncated price model captures the interaction between trading activity, liquidity, and prices in regular market conditions, it does not capture extreme market conditions, such as liquidity crises, extreme volatility, or very rapid changes in liquidity. More precisely, the lower bound $C_L$ on the depth within distance $\delta$ of the best quotes caps the price response per unit of net order flow at $\kappa/C_L$, so that a nearly exhausted top of the book cannot generate arbitrarily large price moves. The upper bound $C_U$ ensures that the price response per unit of order flow never falls below $\kappa/C_U$, so that even in very deep books trading retains a minimal price impact; for $C_U$ large, the resulting distortion is negligible. The clipping of $\Lambda_a$ and $\Lambda_b$, which measure the change in liquidity near the best quotes induced by the autonomous dynamics of the book, bounds their contribution to the mid-price drift uniformly in the state, and the projection $\Pi_R$ caps the instantaneous volatility of the mid-price.

From now on, unless otherwise specified, we consider the truncated version of
$F$ and the corresponding truncated version of the density-price system given by
\eqref{loblim-1}--\eqref{trf}. Since, for each $\mathcal F_T$-measurable input process $g$, the
density-price  \eqref{loblim-1}--\eqref{trf} is uniquely
determined, we occasionally use the shorthand notation
\begin{equation}
UF[g]
=
\left(
u^a[g],u^b[g],F[g]
\right)
\end{equation}
for its solution. We refer to $UF[g]$ as the density-price system determined by $g$, and $F[g]$ denotes the corresponding mid-price process. The argument $[g]$ is omitted whenever no confusion can arise.

\subsection{Optimal Trading Problem Incorporating the Density--Price  System}\label{sec:final-N-player-game}
Combining \Cref{gese} to \Cref{sec:truncated_density_price}, we obtain the
following version of the $N$-player game, with filtration
$\mathcal F^{[N]}=(\mathcal F_t^{[N]})_{t\in[0,T]}$. More precisely, let
\begin{equation}
v_t^{(N)}
=
\frac{1}{N}\sum_{i=1}^N v_t^i.
\end{equation}
The associated density-price system, denoted by $UF\left[v^{(N)}\right]=
\left(
u^a[v^{(N)}],u^b[v^{(N)}],F[v^{(N)}]
\right)$, is then given by \eqref{loblim-1}--\eqref{trf} with $g:=v^{(N)}$.

For each $i=1,\ldots,N$, the inventory, execution price, and cash processes
satisfy 
\begin{align}
dq_t^i
&=
v_t^i\,dt+\sigma^q\,d\beta_t^i, &q^i_0\in \mathbb{R}_{>0},
\\
S_t^i
&=
F_t[v^{(N)}]+av_t^i, &S^i_0\in \mathbb{R}_{>0},
\\
dZ_t^i
&=
-S_t^i\,dq_t^i, &Z^i_0\in \mathbb{R}_{>0}.
\end{align}
Given the strategies
$v^{-i}
=
(v^1,\ldots,v^{i-1},v^{i+1},\ldots,v^N)$
of the other traders, trader $i$ chooses $v^i$ to minimize
\begin{equation}
J_i(v^i,v^{-i})
=
\mathbb{E}
\left[
-Z_T^i
-(q_T^i-r)
\left(
F_T[v^{(N)}]-\xi (q_T^i-r)
\right)
+
\int_0^T
\phi(q_s^i-r)^2\,ds
\right],
\end{equation}
Therefore, the interaction among the traders is transmitted through the
average trading rate $v^{(N)}$. The latter affects the limit order book
densities $u^{a,(N)}$ and $u^{b,(N)}$, which determine the common mid-price
$F^{(N)}$ entering each player's optimization problem. This formulation
provides the $N$-player framework from which the corresponding mean field
game will be derived in the next section.

\section{Mean Field Game of Optimal Trading}\label{FMGg}
We now study the limiting game as the number of agents $N$ tends to infinity, namely the mean field game associated with
the $N$-player game introduced in \Cref{sec:final-N-player-game}.
Writing $W^0=(W^a,W^b)$, we denote by
$\mathcal F^{W^{0}}=(\mathcal F_t^{W^{0}})_{t\in[0,T]}$ the filtration
given by
\begin{equation}
\mathcal F_t^{W^{0}}
=
\sigma\left(
W_s^0:0\leq s\leq t
\right)\vee\mathcal N,
\qquad t\in[0,T].
\end{equation}
For simplicity, we write $\mathcal M_{W^0}$ for
$\mathcal M_{\mathcal F^{W^{0}}}(0,T)$. The mean field game is characterized by the solution of the following limiting system. In the limiting system, as the number of traders $N$ goes to infinity, the empirical average control $v^{(N)}$ is replaced
by a process $\bar{v}\in\mathcal M_{W^0}$. The density-price system, denoted by $UF[\bar{v}]$, is therefore given by \eqref{loblim-1}--\eqref{trf} with $g:=\bar{v}$.

For each representative trader $i$, let $\mathcal F^{W^{0},\beta^i}=(\mathcal F_t^{W^{0},\beta^i})_{t\in[0,T]}$ be her information set, where 
\begin{equation}
\mathcal F_t^{W^{0},\beta^i}
=
\sigma\left(
W_s^a,W_s^b,\beta_s^i:0\leq s\leq t
\right)\vee\mathcal N,
\qquad t\in[0,T].
\end{equation}
The admissible control $v^i$ is taken from
$\mathcal M_{\mathcal F^{W^{0},\beta^i}}(0,T)$, and the corresponding inventory,
execution price, and cash processes satisfy
\begin{align} \label{limsys}
dq_t^i
&=
v_t^i\,dt+\sigma^q\,d\beta_t^i,&q^i_0\in \mathbb{R}_{>0},
\\
S_t^i
&=
F_t[\bar{v}]+av_t^i,&S^i_0\in \mathbb{R}_{>0},
\\
dZ_t^i
&=
-S_t^i\,dq_t^i, &Z^i_0\in \mathbb{R}_{>0},
\end{align}
Moreover, the cost functional is given by 
\begin{equation}\label{limcost}
J^\infty(v^i;\bar{v})
=
\mathbb{E}
\left[
-Z_T^i
-(q_T^i-r)
\left(
F_T[\bar{v}]-\xi (q_T^i-r
\right)
+
\int_0^T
\phi(q_s^i-r)^2\,ds
\right].
\end{equation}
The mean field equilibrium is characterized in two steps:
\begin{itemize}
\item[(i)] \emph{Stochastic optimal control.} Treating $\bar{v}$ as a given input $g$, solve the stochastic control problem of a representative trader $i$ to obtain the optimal strategy $v^{i,*}$.
\item[(ii)] \emph{Mean field consistency condition.} Require that
\begin{equation}
\bar{v}_t = \mathbb{E}\left[v_t^{i,*}\mid\mathcal{F}_t^{W^{0}}\right].
\end{equation}
\end{itemize}
Thus, at equilibrium, the mean field $\bar{v}$ coincides with the conditional expectation of the representative trader's optimal strategy given the common noise. In particular, when all traders follow their equilibrium strategies,
the empirical average of their controls converges to the mean field,
i.e.,
\begin{equation}
\frac{1}{N}\sum_{i=1}^N v_i(t)
\stackrel{\mathrm{q.m.}}{\longrightarrow}
\bar{v}(t).
\end{equation}
\subsection{Stochastic Optimal Control of Representative Trader}\label{sincon}
In this section, we address the control problem of a representative trader and drop the index $i$, since the traders are homogeneous. Fix $g\in\mathcal M_{W^0}$. The corresponding density-price system $UF[g]=\left(
u^a[g],u^b[g],F[g]\right)$ is then given by \eqref{loblim-1}--\eqref{trf}, and we omit the argument $[g]$ throughout this section. Given this
density-price system, the representative agent seeks to minimize the cost
functional
\begin{equation}
J(v;g)
=
\mathbb{E}
\left[
-Z_T
-(q_T-r)\left(F_T-\xi (q_T-r)\right)
+
\int_0^T\phi (q_s-r)^2\,ds
\right],
\end{equation}
subject to the dynamics 
\begin{equation}
S_t=F_t+av_t,
\qquad
dZ_t=-S_t\,dq_t,
\qquad
dq_t=v_t\,dt+\sigma^q\,d\beta_t,
\end{equation}
with initial conditions $S_0, Z_0, q_0 \in \mathbb{R}_{>0}$.
It follows that
\begin{equation}
Z_T=
Z_0-\int_0^T S_t\,dq_t
=
Z_0
-\int_0^T
\left(F_t+av_t\right)v_t\,dt
-\sigma^q\int_0^T
\left(F_t+av_t\right)\,d\beta_t.
\end{equation}
Under the admissibility and integrability conditions imposed on $v$ and
$F$, the stochastic integral has zero expectation. Therefore,
\begin{align}
J(v;g)
&=
-Z_0
+
\mathbb{E}
\left[
\int_0^T
\left(
F_tv_t
+
av_t^2
+
\phi (q_t-r)^2
\right)dt
-(q_T-r)F_T
+
\xi (q_T-r)^2
\right].
\end{align}
Since $Z_0$ is independent of the control, it can be omitted from the
optimization problem. Hence, for a fixed $g$, the optimal control problem
is equivalently given by
\begin{equation}\label{cost-rep}
J(v) = 
\mathbb{E}
\left[
\int_0^T
\left(
F_tv_t
+
av_t^2
+
\phi (q_t-r)^2
\right)dt
-(q_T-r)F_T
+
\xi (q_T-r)^2
\right],
\end{equation}
subject to 
\begin{equation}\label{dyn-rep}
dq_t=v_t\,dt+\sigma^q\,d\beta_t,
\qquad q_0\in\mathbb R_{>0}.
\end{equation}
\begin{theorem}
Let $a>0$, and consider the optimal control problem defined by
\eqref{dyn-rep} and \eqref{cost-rep}.

Let $P$ solve the Riccati equation
\begin{equation}\label{eq_P}
\dot P_t
=
\frac{P_t^2}{2a}-2\phi,
\qquad
P_T=2\xi,
\end{equation}
and let $(\varphi,Z)$ solve
\begin{equation}\label{eq_varphi}
d\varphi_t
=
\frac{P_t}{2a}
(F_t+\varphi_t)\,dt
+
Z_t\,dW^{0}_t,
\qquad
\varphi_T=-F_T.
\end{equation}
Then the unique optimal control, $v_t^*$, $t\in[0,T]$, is given by
\begin{equation}
v_t^*
=
-\frac{
F_t+P_t(q_t^*-r)+\varphi_t
}{
2a
}.
\end{equation}
\end{theorem}

\begin{proof}
Consider a perturbed control $v^\epsilon=v+\epsilon\omega$ with $\omega \in \mathcal M_{\mathcal F^{W^{0},\beta}}(0,T)$ adapted. The state under $v$ and under $v^\epsilon$ is
\begin{equation}
q_t = q_0+\int_0^t v_s\,ds+\int_0^t\sigma^q\,d\beta_s,\qquad
q_t^\epsilon = q_0+\int_0^t (v_s+\epsilon\omega_s)\,ds+\int_0^t\sigma^q\,d\beta_s .
\end{equation}
Hence, with $Q_t:=\int_0^t\omega_s\,ds$, we have
\begin{equation}
q^\epsilon_t=q_t+\epsilon Q_t .
\end{equation}
 The difference of the corresponding costs is therefore given by
\begin{align}
J(v^\epsilon)-J(v)
&=\epsilon\,\mathbb E\Big[\int_0^T\!\big(F_t\omega_t+2a\,v_t\omega_t+2\phi\,Q_t(q_t-r)\big)dt-Q_TF_T+2\xi\,Q_T(q_T-r)\Big]\nonumber\\
&\quad+\epsilon^2\,\mathbb E\Big[\int_0^T\!\big(a\,\omega_t^2+\phi\,Q_t^2\big)dt+\xi\,Q_T^2\Big].
\label{eq:cost-diff}
\end{align}
 Using Fubini, $\int_0^T Q_t\,g_t\,dt=\int_0^T\omega_s\int_s^T g_t\,dt\,ds$, which gives
\begin{align}
J(v^\epsilon)-J(v)
&=\epsilon\,\mathbb E\!\int_0^T\omega_s\Big[F_s+2a\,v_s+2\phi\!\int_s^T(q_t-r)\,dt+2\xi(q_T-r)-F_T\Big]ds\nonumber\\
&\quad+\epsilon^2\,\mathbb E\Big[\int_0^T\!\big(a\,\omega_t^2+\phi\,Q_t^2\big)dt+\xi\,Q_T^2\Big].
\label{eq:cost-diff-omega}
\end{align}
Consequently, the G\^ateaux derivative at $v$ in the direction $\omega$ is
\begin{equation}
\frac{d}{d\epsilon}J(v^\epsilon)\Big|_{\epsilon=0}
= \mathbb{E}\!\int_0^T \omega_s
\Big[ F_s + 2a\,v_s + 2\phi\!\int_s^T (q_t - r)\,dt + 2\xi\,(q_T - r) - F_T \Big]\,ds,
\end{equation}
Since $J$ is strictly convex, $v^*$ is optimal if and only if
\begin{equation}
F_s+2a\,v_s^*+\mathbb E\Big[2\phi\!\int_s^T(q_t^*-r)\,dt+2\xi(q_T^*-r)-F_T\,\Big|\,\mathcal F_s\Big]=0,
\qquad ds\otimes d\mathbb P\text{-a.e.}
\end{equation}
Splitting $\int_s^T=\int_0^T-\int_0^s$, the above condition may be written as
\begin{equation}
F_s+2a\,v_s^*+2\phi\Big(\mathbb E\Big[\int_0^T(q_t^*-r)\,dt\,\Big|\,\mathcal F_s\Big]-\int_0^s(q_t^*-r)\,dt\Big)
+2\xi\,\mathbb E\big[q_T^*-r\mid\mathcal F_s\big]-\mathbb E[F_T\mid\mathcal F_s]=0 .
\end{equation}

Then the conditional expectations are naturally collected into a martingale given by
\begin{equation}
M_t:=\mathbb E\Big[F_T-2\phi\int_0^T(q_u^*-r)\,du-2\xi(q_T^*-r)\,\Big|\,\mathcal F_t\Big],\qquad t\in[0,T].
\end{equation}
Solving the optimality condition for $v$ gives
\begin{equation}
v_t^*=\frac1{2a}\Big[M_t-F_t+2\phi\int_0^t(q_u^*-r)\,du\Big],\qquad dt\otimes d\mathbb P\text{-a.e.}
\end{equation}
By the martingale representation theorem, $dM_t=\zeta^\beta_t\,d\beta_t+\zeta^0_t\,dW^0_t$, hence differentiating this formula yields an SDE for $v$ given by
\begin{equation}
dv_t^*=\frac1{2a}\Big[\,dM_t-dF_t+2\phi\,(q_t^*-r)\,dt\Big],\qquad
dM_t=\zeta^\beta_t\,d\beta_t+\zeta^0_t\,dW^0_t. \tag{A}
\end{equation}

We now look for the optimal control in feedback form by adopting the ansatz
\begin{equation}
v_t^*=-\frac{F_t+P_t(q_t^*-r)+\varphi_t}{2a},
\end{equation}
where $P$ is deterministic and $\varphi$ is an It\^o process. Subsequently, differentiating the ansatz the SDE 
\begin{equation}
dv_t^*=-\frac1{2a}\Big[dF_t+\big(\dot P_t(q_t^*-r)+P_tv_t^*\big)dt+P_t\sigma^q\,d\beta_t+d\varphi_t\Big]. \tag{B}
\end{equation}
Substituting the ansatz again into the term $P_tv_t$ and collecting the terms in $q_t-r$, we obtain
\begin{equation}
dv_t^*=-\frac1{2a}\Big[dF_t+\Big(\Big(\dot P_t-\frac{P_t^2}{2a}\Big)(q_t^*-r)-\frac{P_t}{2a}\big(F_t+\varphi_t\big)\Big)dt+P_t\sigma^q\,d\beta_t+d\varphi_t\Big]. \tag{B$'$}
\end{equation}

Equating (A) and (B$'$), the $dF_t$ terms cancel, we obtain \eqref{eq_P}  and \eqref{eq_varphi}, where 
$Z_t:=-\zeta^0_t$.
\end{proof}
\subsection{Fixed-Point Problem and Nash Equilibrium}\label{sec:fixed_point_equilibrium}

In this section, we study the mean field consistency condition as a fixed point problem. We then establish the existence and
uniqueness of a fixed point for a suitably truncated system under a
small-time condition. The consistency condition is given by
\begin{equation}
g_t
=
\mathbb{E}\left[
v_t^* \mid \mathcal{F}_t^{W^{0}}
\right],
\qquad t\in[0,T].
\label{consistency}
\end{equation}
Using the expression for $v^*$, the optimal inventory satisfies
\begin{equation}\label{q_optimal}
dq_t^*
=
-\frac{P_t}{2a}(q_t^*-r)\,dt
-\frac{F_t+\varphi_t}{2a}\,dt
+\sigma^q\,d\beta_t.
\end{equation}
Since $q_0$, $\beta$, and $W^{0}$ are mutually independent, we may decompose
\begin{equation}
q_t^*=q_t^{W^{0}}+q_t^\perp,
\end{equation}
where 
\begin{align}
dq_t^{W^{0}}
&=
-\frac{P_t}{2a}(q_t^{W^{0}}-r)\,dt
-\frac{F_t+\varphi_t}{2a}\,dt,
&q_0^{W^{0}}=\mathbb E[q_0],
\\
dq_t^\perp
&=
-\frac{P_t}{2a}q_t^\perp\,dt
+\sigma^q\,d\beta_t,&
q_0^\perp=q_0-\mathbb E[q_0].
\end{align}
In particular, $q^{W^{0}}$ is $\mathcal F^{W^{0}}$-adapted, while $q^\perp$ is
independent of $W^{0}$, and
\begin{equation}
\mathbb E[q_t^*\mid\mathcal F_t^{W^{0}}]=q_t^{W^{0}}.
\end{equation}
The consistency condition therefore becomes
\begin{equation}
g_t
=
-\frac{
F_t+P_t(q_t^{W^{0}}-r)+\varphi_t
}{
2a
},
\qquad t\in[0,T].
\end{equation}

Here, for a given process $g$, the density-price system $UF[g]$ is given by \eqref{loblim-1}--\eqref{trf}, and the resulting processes 
$q^{W^{0}}[g]$ and $\varphi[g]$ are then determined,
with $g$ treated as an input.
This naturally defines the mapping
$\Psi:\mathcal M_{W^0}\to\mathcal M_{W^0} $ given by
\begin{equation}
\Psi(g)_t
=
-\frac{
F_t[g]+P_t(q_t^{W^{0}}[g]-r)+\varphi_t[g]
}{
2a
},
\qquad t\in[0,T].
\label{fixedpointmap}
\end{equation}
Hence, the consistency condition is equivalent to the fixed point condition
\begin{equation}
g=\Psi(g).
\end{equation}

In other words, a solution to the mean field consistency condition is characterized by a
solution to the system given by
\begin{equation} 
\label{bwpt}
\left\{
\begin{aligned}
&UF[g]
=
\left(
u^a[g],u^b[g],F[g]
\right),
\\
dq_t^{W^{0}}
&=
-\frac{P_t}{2a}(q_t^{W^{0}}-r)\,dt
-\frac{F_t+\varphi_t}{2a}\,dt,
\qquad
q_0^{W^{0}}=\mathbb{E}[q_0],
\\
d\varphi_t
&=
\frac{P_t}{2a}
\left(F_t+\varphi_t\right)dt
+
Z_t\,dW^{0}_t,
\qquad\,\,\qquad
\varphi_T=-F_T,
\\
g_t
&=
-\frac{
F_t+P_t(q_t^{W^{0}}-r)+\varphi_t
}{
2a}.
\end{aligned}
\right.
\end{equation}
We first impose the following assumptions on the coefficients of the limit order
book dynamics, adapted from \Cref{ass:regularity}.

\begin{assumption}
\label{asmfg}
Let $\frac14<\alpha<\frac12$. The following conditions hold:
\begin{itemize}
    \item[(i)] The initial conditions satisfy $u^a_{0}\in D_{A_a}\left(\alpha+\frac12,2\right)$ and $u^b_{0}\in D_{A_b }\left(\alpha+\frac12,2\right)$.

\item[(ii)] The mappings
\begin{equation*}
G^a:\mathbb{R}\times H^a\times[0,T]
\longrightarrow
D_{A_a}\left(\alpha+\frac12,2\right),
\qquad
G^b:\mathbb{R}\times H^b\times[0,T]
\longrightarrow
D_{A_b }\left(\alpha+\frac12,2\right),
\end{equation*}
and
\begin{equation*}
J^a:H^a\times[0,T]
\longrightarrow
D_{A_a}\left(\alpha+\frac12,2\right),
\qquad
J^b:H^b\times[0,T]
\longrightarrow
D_{A_b }\left(\alpha+\frac12,2\right),
\end{equation*}
are Borel measurable and bounded by a positive constant $K$.
Moreover, there exists a constant $L>0$ such that, for all $t\in[0,T]$,
\begin{align*}
\|G^a(v,u,t)-G^a(\tilde v,\tilde u,t)\|_
{D_{A_a}\left(\alpha+\frac12,2\right)}
&\leq
L\left(
|v-\tilde v|
+
\|u-\tilde u\|_{H^a}
\right),
\\
\|G^b(v,u,t)-G^b(\tilde v,\tilde u,t)\|_
{D_{A_b }\left(\alpha+\frac12,2\right)}
&\leq
L\left(
|v-\tilde v|
+
\|u-\tilde u\|_{H^b}
\right),
\\
\|J^a(u,t)-J^a(\tilde u,t)\|_
{D_{A_a}\left(\alpha+\frac12,2\right)}
&\leq
L\|u-\tilde u\|_{H^a},
\\
\|J^b(u,t)-J^b(\tilde u,t)\|_
{D_{A_b }\left(\alpha+\frac12,2\right)}
&\leq
L\|u-\tilde u\|_{H^b}.
\end{align*}

\item[(iii)] For the mappings
\begin{equation*}
B^a\in
\mathcal L\left(
H^a,
L_2\left(H_0^a;D_{A_a}(\alpha,2)\right)
\right),
\qquad
B^b\in
\mathcal L\left(
H^b,
L_2\left(H_0^b;D_{A_b }(\alpha,2)\right)
\right),
\end{equation*}
there exist constants $\varrho, \rho>0$ such that
\begin{equation*}
\begin{aligned}
\|B^a(x)\|_{L_2(U_0^a;D_{A_a}(\alpha,2))}^2
&\leq \varrho \|x\|_{H^a}^2
\leq \rho\|x\|_{D_{A_a}(\alpha+\frac12,2)}^2,\\
\|B^b(x)\|_{L_2(U_0^b;D_{A_b }(\alpha,2))}^2
&\leq \varrho \|x\|_{H^b}^2
\leq \rho\|x\|_{D_{A_b }(\alpha+\frac12,2)}^2,
\end{aligned}
\end{equation*}
with $\rho<1-2\alpha$.

\end{itemize}
\end{assumption}

Under the above assumption, for every $g\in\mathcal M_{W^0}$, \Cref{prop:regularity} ensures that the system \eqref{loblim-1}-\eqref{loblim-2}
admits a unique mild solution satisfying 
\begin{equation}
u^a\in
\mathcal M_{W^0}\left(
0,T;
D_{A_a}\left(\alpha+\frac12,2\right)
\right),
\qquad
u^b\in
\mathcal M_{W^0}\left(
0,T;
D_{A_b }\left(\alpha+\frac12,2\right)
\right).
\end{equation}
For simplicity, we use the notation $h^a:=\mathbf{1}_{[0,\delta]}$ and
$h^b:=\mathbf{1}_{[-\delta,0]}$. For the price dynamics, we further introduce the notation given by 
\begin{align}
\label{Xi_2}
\Xi_t^b
&=
\int_0^t
\frac{
T_{-C_U,C_U}\Lambda_b(u_s^b)
+
\langle h^b,J_s^b\rangle
+
\langle h^b,G^b(g_s,u_s^b)\rangle
}{
T_{C_L,C_U}\left(-\langle h^b,u_s^b\rangle\right)
}\,ds
+
\int_0^t
\Pi_R^b\!\left(
\frac{
B^b(u_s^b)^*h^b
}{
T_{C_L,C_U}\left(-\langle h^b,u_s^b\rangle\right)
}
\right)dW_s^b,
\\
\Xi_t^a
&=
\int_0^t
\frac{
T_{-C_U,C_U}\Lambda_a(u_s^a)
+
\langle h^a,J_s^a\rangle
+
\langle h^a,G^a(g_s,u_s^a)\rangle
}{
T_{C_L,C_U}\langle h^a,u_s^a\rangle
}\,ds
+
\int_0^t
\Pi_R^a\!\left(
\frac{
B^a(u_s^a)^*h^a
}{
T_{C_L,C_U}\langle h^a,u_s^a\rangle
}
\right)dW_s^a.
\label{Xi_1}
\end{align}
Then the truncated mid-price dynamics can be written as
\begin{equation}\label{tr3}
dF_t
=
-\kappa\left(
d\Xi_t^b+d\Xi_t^a
\right).
\end{equation}

The mean field consistency condition is therefore characterized by the coupled system 
\begin{equation}
\label{trsystem}
\left\{
\begin{aligned}
u_t^a
&=
S_a(t)u^a_0
+
\int_0^t
S_a(t-s)
\left[
G^a(g_s,u_s^a,s)
+
J^a(u_s^a,s)
\right]ds
+
\int_0^t
S_a(t-s)B^a(u_s^a)\,dW_s^a,
\\
u_t^b
&=
S_b(t)u^b_0
+
\int_0^t
S_b(t-s)
\left[
G^b(g_s,u_s^b,s)
+
J^b(u_s^b,s)
\right]ds
+
\int_0^t
S_b(t-s)B^b(u_s^b)\,dW_s^b,
\\
F_t
&=
F_0-
\kappa\left(\Xi_t^a+\Xi_t^b\right),
\qquad F_0\in\mathbb R_{>0},
\\
dq_t^{W^{0}}
&=
-\frac{P_t}{2a}(q_t^{W^{0}}-r)\,dt
-\frac{F_t+\varphi_t}{2a}\,dt,
\qquad q_0^{W^{0}}=\mathbb E[q_0],
\\
d\varphi_t
&=
\frac{P_t}{2a}
\left(F_t+\varphi_t\right)dt
+
Z_t\,dW^{0}_t,
\qquad\varphi_T=-F_T,
\\
g_t
&=
-\frac{
F_t+P_t(q_t^{W^{0}}-r)+\varphi_t
}{
2a}.
\end{aligned}
\right.
\end{equation}
where 
\begin{equation}
\dot P_t
=
\frac{P_t^2}{2a}-2\phi,
\qquad
P_T=2\xi.\label{P_eq}
\end{equation}
To establish the existence of a fixed point, and hence a solution to the above
system, we prove that the mapping $\Psi:\mathcal M_{W^0}\to\mathcal M_{W^0}$ given by \eqref{fixedpointmap} is a contraction. To this end, we first derive a sequence of stability
estimates with respect to the input process $g$. We begin with the dependence
of the limit order book densities on $g$.

\begin{lemma}
\label{lem:lob_lipschitz_g}
Suppose that \Cref{asmfg} holds and that
\begin{equation}
\label{bound1}
\sqrt{5}\,M_A\,L\,T
<
1-
\sqrt{\frac{\rho}{1-2\alpha}}.
\end{equation}
Let $g^1,g^2\in\mathcal M_{W^0}$, and, for $\ell\in\{a,b\}$, let
$u^{\ell,1}$ and $u^{\ell,2}$ be the corresponding mild solutions of
\eqref{loblim-1}--\eqref{loblim-2}. Then the following estimate holds: 
\begin{equation}
\|u^{\ell,1}-u^{\ell,2}\|_
{\mathcal M_{W^0}(0,T;D_{A_\ell})}^2
\leq
C_1(T)
|g^1-g^2|_{\mathcal M_{W^0}(0,T)}^2,
\qquad
\ell\in\{a,b\},
\end{equation}
where
\begin{equation}
C_1(T)
=
\frac{
5M_A^2L^2T^2
}{
\left(
1-\sqrt{\dfrac{\rho}{1-2\alpha}}
\right)^2
-
5M_A^2L^2T^2
}.
\end{equation}
\end{lemma}

\begin{proof}
Fix $\ell\in\{a,b\}$ and set
\begin{equation}
\Delta u
=
u^{\ell,1}-u^{\ell,2},
\qquad
\Delta g
=
g^1-g^2.
\end{equation}
Then
\begin{equation}
\Delta u
=
\Delta u^D+\Delta u^M,
\end{equation}
where
\begin{align}
\Delta u_t^D
&=
\int_0^t
S_\ell(t-s)
\Big[
G^\ell(g_s^1,u_s^{\ell,1},s)
-
G^\ell(g_s^2,u_s^{\ell,2},s)
+
J^\ell(u_s^{\ell,1},s)
-
J^\ell(u_s^{\ell,2},s)
\Big]ds,
\\
\Delta u_t^M
&=
\int_0^t
S_\ell(t-s)
B_\ell(\Delta u_s)\,dW_s^\ell.
\end{align}
By \Cref{asmfg},
\begin{align}
&\left\|
G^\ell(g_s^1,u_s^{\ell,1},s)
-
G^\ell(g_s^2,u_s^{\ell,2},s)
+
J^\ell(u_s^{\ell,1},s)
-
J^\ell(u_s^{\ell,2},s)
\right\|_{D_{A_\ell}}
\leq
L|\Delta g_s|
+
2L\|\Delta u_s\|_{D_{A_\ell}}.
\end{align}
Hence,
\begin{align}
\|\Delta u^D\|_{\mathcal M_{W^0}(0,T;D_{A_\ell})}^2
&\leq
5M_A^2L^2T^2
\left(
|\Delta g|_{\mathcal M_{W^0}(0,T)}^2
+
\|\Delta u\|_{\mathcal M_{W^0}(0,T;D_{A_\ell})}^2
\right).
\end{align}
Moreover, by the stochastic convolution estimate used in
Proposition~\ref{prop:regularity} and \Cref{asmfg},
\begin{equation}
\|\Delta u^M\|_{\mathcal M_{W^0}(0,T;D_{A_\ell})}^2
\leq
\frac{\rho}{1-2\alpha}
\|\Delta u\|_{\mathcal M_{W^0}(0,T;D_{A_\ell})}^2.
\end{equation} 
Therefore, for every $\varepsilon>0$,
\begin{align}
\|\Delta u\|_{\mathcal M_{W^0}(0,T;D_{A_\ell})}^2
&\leq
\left(1+\frac1\varepsilon\right)
5M_A^2L^2T^2
|\Delta g|_{\mathcal M_{W^0}(0,T)}^2
\nonumber\\
&\quad+
\left[
\left(1+\frac1\varepsilon\right)
5M_A^2L^2T^2
+
(1+\varepsilon)
\frac{\rho}{1-2\alpha}
\right]
\|\Delta u\|_{\mathcal M_{W^0}(0,T;D_{A_\ell})}^2.
\end{align}

We first note that, under \eqref{bound1}, there exists a nonempty range of
$\varepsilon>0$ such that
\begin{equation}
\left(1+\frac1\varepsilon\right)
5M_A^2L^2T^2
+
(1+\varepsilon)
\frac{\rho}{1-2\alpha}
<1.
\end{equation}
For every $\varepsilon$ in this feasible range, rearranging the preceding
estimate gives
\begin{equation}
\|\Delta u\|_{\mathcal M_{W^0}(0,T;D_{A_\ell})}^2
\leq
\frac{
\left(1+\frac1\varepsilon\right)5M_A^2L^2T^2
}{
1-
\left(1+\frac1\varepsilon\right)5M_A^2L^2T^2
-
(1+\varepsilon)\dfrac{\rho}{1-2\alpha}
}
|\Delta g|_{\mathcal M_{W^0}(0,T)}^2.
\end{equation}

We now minimize the coefficient on the right-hand side, viewed as a
function of $\varepsilon$, over the feasible range. The minimum is attained
at
\begin{equation}
\varepsilon^*
=
\frac{
1-\sqrt{\dfrac{\rho}{1-2\alpha}}
}{
\sqrt{\dfrac{\rho}{1-2\alpha}}
},
\end{equation}
and the corresponding minimum value is $C_1(T)$. Consequently,
\begin{equation}
\|\Delta u\|_{\mathcal M_{W^0}(0,T;D_{A_\ell})}^2
\leq
C_1(T)
|\Delta g|_{\mathcal M_{W^0}(0,T)}^2.
\end{equation}
This proves the desired estimate.
\end{proof}

The preceding lemma provides the required stability of the limit order book
densities. We next use this estimate to control the corresponding ask- and
bid-side components of the truncated mid-price process.

\begin{lemma}
\label{lem:Xi_lipschitz}
Suppose that \Cref{asmfg} and the condition \eqref{bound1} hold. Let
$g^1,g^2\in\mathcal M_{W^0}$, and, for $\ell\in\{a,b\}$, let
$u^{\ell,1}$ and $u^{\ell,2}$ be the corresponding mild solutions of
\eqref{loblim-1}--\eqref{loblim-2}. Let $\Xi^{\ell,1}$ and $\Xi^{\ell,2}$ be the corresponding
processes defined in \eqref{Xi_1} and \eqref{Xi_2}. Then, for every $t\in[0,T]$,
\begin{equation}
\mathbb E
\left|
\Xi_t^{\ell,1}-\Xi_t^{\ell,2}
\right|^2
\leq
C_{\Xi,\ell}(T)
|g^1-g^2|_{\mathcal M_{W^0}(0,T)}^2,
\qquad
\ell\in\{a,b\},
\end{equation}
where
\begin{equation}
C_{\Xi,\ell}(T)
=
4T c_{\ell,g}^2
+
\left(
4T c_{\ell,u}^2
+
2c_{\ell,\sigma}^2
\right)C_1(T),
\end{equation}
with $C_1(T)$ given in  \Cref{lem:lob_lipschitz_g}, and
\begin{align}
c_{\ell,g}
&=
\frac{\sqrt{\delta}L}{C_L},
\\
c_{\ell,u}
&=
\frac{
\|\Lambda_\ell\|_{D_{A_\ell}^{'}}
+
2\sqrt{\frac{\rho}{\varrho}}\sqrt{\delta}L
}{C_L}
+
\frac{
K_{\varsigma}^\ell\sqrt{\frac{\rho}{\varrho}}\sqrt{\delta}
}{C_L^2},
\\
c_{\ell,\sigma}
&=
\frac{\sqrt{\frac{\rho}{\varrho}}}{C_L}
\left[
\|h^\ell B_\ell\|_{
\mathcal L\left(
H^\ell,L_2(H_0^\ell;\mathbb R)
\right)}
+
R\sqrt{\delta}
\right],\\
K_{\varsigma}^\ell
&:=
C_U+2K\sqrt{\delta}.
\end{align}
\end{lemma}

\begin{proof}
Fix $\ell\in\{a,b\}$ and set
\begin{equation}
\Delta g_t:=g_t^1-g_t^2,
\qquad
\Delta u_t^\ell:=u_t^{\ell,1}-u_t^{\ell,2}.
\end{equation}
For $i=1,2$, define
\begin{equation}
\varsigma_t^{\ell,i}
:=
T_{-C_U,C_U}\Lambda_\ell(u_t^{\ell,i})
+
\varsigma_\ell
\left[
\langle h^\ell,J^\ell(u_t^{\ell,i},t)\rangle_{H^\ell}
+
\langle h^\ell,G^\ell(g_t^i,u_t^{\ell,i},t)\rangle_{H^\ell}
\right],
\end{equation}
where $\varsigma_a=1$ and $\varsigma_b=-1$.

Since $\Lambda_\ell$ is a bounded linear functional on $D_{A_\ell}$,
\begin{equation}
\left|
\Lambda_\ell(u_t^{\ell,1})
-
\Lambda_\ell(u_t^{\ell,2})
\right|
\leq
\|\Lambda_\ell\|_{D_{A_\ell}^{'}}
\|\Delta u_t^\ell\|_{D_{A_\ell}}.
\end{equation}
Since $T_{-C_U,C_U}$ is $1$-Lipschitz and
\begin{equation}
\|h^a\|_{H^a}
=
\|h^b\|_{H^b}
=
\sqrt{\delta},
\end{equation}
Assumption~\ref{asmfg} gives
\begin{align}
\left|
\varsigma_t^{\ell,1}
-
\varsigma_t^{\ell,2}
\right|
\leq
\sqrt{\delta}L|\Delta g_t|
+
\left(
\|\Lambda_\ell\|_{D_{A_\ell}^{'}}
+
2\sqrt{\frac{\rho}{\varrho}}\sqrt{\delta}L
\right)
\|\Delta u_t^\ell\|_{D_{A_\ell}}.
\end{align}
Moreover,
\begin{equation}
|\varsigma_t^{\ell,i}|
\leq
K_{\varsigma}^\ell,
\qquad
i=1,2.
\end{equation}

Since
\begin{equation}
T_{C_L,C_U}
\langle h^\ell,u_t^{\ell,i}\rangle_{H^\ell}
\geq C_L
\end{equation}
and $T_{C_L,C_U}$ is $1$-Lipschitz, we obtain
\begin{align}
&
\left|
\frac{
\varsigma_t^{\ell,1}
}{
T_{C_L,C_U}
\langle h^\ell,u_t^{\ell,1}\rangle_{H^\ell}
}
-
\frac{
\varsigma_t^{\ell,2}
}{
T_{C_L,C_U}
\langle h^\ell,u_t^{\ell,2}\rangle_{H^\ell}
}
\right|
\leq
c_{\ell,g}|\Delta g_t|
+
c_{\ell,u}
\|\Delta u_t^\ell\|_{D_{A_\ell}}.
\end{align}

For the diffusion coefficient, set
\begin{equation}
d_t^i
:=
T_{C_L,C_U}
\langle h^\ell,u_t^{\ell,i}\rangle_{H^\ell},
\qquad
i=1,2.
\end{equation}
Using the $1$-Lipschitz property of $\Pi_R^\ell$, we obtain
\begin{align}
&
\left\|
\Pi_R^\ell\left(
\frac{
h^\ell B_\ell(u_t^{\ell,1})
}{
d_t^1
}
\right)
-
\Pi_R^\ell\left(
\frac{
h^\ell B_\ell(u_t^{\ell,2})
}{
d_t^1
}
\right)
\right\|_{L_2(H_0^\ell;\mathbb R)}
\leq
\frac{
\sqrt{\frac{\rho}{\varrho}}
\|h^\ell B_\ell\|_{
\mathcal L\left(
H^\ell,L_2(H_0^\ell;\mathbb R)
\right)}
}{
C_L
}
\|\Delta u_t^\ell\|_{D_{A_\ell}}.
\end{align}
Moreover,
\begin{align}
\left\|
\Pi_R^\ell\left(
\frac{
h^\ell B_\ell(u_t^{\ell,2})
}{
d_t^1
}
\right)
-
\Pi_R^\ell\left(
\frac{
h^\ell B_\ell(u_t^{\ell,2})
}{
d_t^2
}
\right)
\right\|_{L_2(H_0^\ell;\mathbb R)}
\leq
\frac{R}{C_L}
|d_t^1-d_t^2|
\leq
\frac{R\sqrt{\frac{\rho}{\varrho}}\sqrt{\delta}}{C_L}
\|\Delta u_t^\ell\|_{D_{A_\ell}}.
\end{align}
Consequently,
\begin{align}
&
\left\|
\Pi_R^\ell\left(
\frac{
h^\ell B_\ell(u_t^{\ell,1})
}{
T_{C_L,C_U}
\langle h^\ell,u_t^{\ell,1}\rangle_{H^\ell}
}
\right)
-
\Pi_R^\ell\left(
\frac{
h^\ell B_\ell(u_t^{\ell,2})
}{
T_{C_L,C_U}
\langle h^\ell,u_t^{\ell,2}\rangle_{H^\ell}
}
\right)
\right\|_{L_2(H_0^\ell;\mathbb R)}
\leq
c_{\ell,\sigma}
\|\Delta u_t^\ell\|_{D_{A_\ell}}.
\end{align}

Using the Cauchy--Schwarz inequality and the It\^o isometry, we obtain
\begin{align}
\mathbb E
\left|
\Xi_t^{\ell,1}-\Xi_t^{\ell,2}
\right|^2
&\leq
4t c_{\ell,g}^2
\mathbb E\int_0^t
|\Delta g_s|^2\,ds
+
\left(
4t c_{\ell,u}^2
+
2c_{\ell,\sigma}^2
\right)
\mathbb E\int_0^t
\|\Delta u_s^\ell\|_{D_{A_\ell}}^2\,ds.
\end{align}
Applying Lemma~\ref{lem:lob_lipschitz_g} and using $t\leq T$ yields
\begin{equation}
\mathbb E
\left|
\Xi_t^{\ell,1}-\Xi_t^{\ell,2}
\right|^2
\leq
C_{\Xi,\ell}(T)
|g^1-g^2|_{\mathcal M_{W^0}(0,T)}^2.
\end{equation}
This proves the desired estimate.
\end{proof}

Combining the estimates for the ask and bid sides yields the following
stability estimate for the truncated mid-price process.

\begin{proposition}
\label{prop:Fstability}
Let $g^1,g^2\in\mathcal M_{W^0}$, and let $F^1,F^2$ be the corresponding
mid-price processes given in \eqref{tr3}. Then, for every $t\in[0,T]$,
\begin{equation}
\mathbb E|F_t^1-F_t^2|^2
\leq
C_F(T)
\mathbb E\int_0^T
|g_s^1-g_s^2|^2\,ds,
\label{Fstability}
\end{equation}
where
\begin{equation}
C_F(T)
:=
2\kappa^2
\left(
C_{\Xi,a}(T)+C_{\Xi,b}(T)
\right).
\label{CF}
\end{equation}
\end{proposition}

\begin{proof}
Since
\begin{equation}
F_t^1-F_t^2
=
\kappa
\left[
(\Xi_t^{a,1}-\Xi_t^{a,2})
-
(\Xi_t^{b,1}-\Xi_t^{b,2})
\right],
\end{equation}
we have
\begin{align}
\mathbb E|F_t^1-F_t^2|^2
&\leq
2\kappa^2
\mathbb E|\Xi_t^{a,1}-\Xi_t^{a,2}|^2
+
2\kappa^2
\mathbb E|\Xi_t^{b,1}-\Xi_t^{b,2}|^2
\nonumber\\
&\leq
2\kappa^2
\left(
C_{\Xi,a}(T)+C_{\Xi,b}(T)
\right)
\mathbb E\int_0^T
|g_s^1-g_s^2|^2\,ds
\nonumber\\
&=
C_F(T)
\mathbb E\int_0^T
|g_s^1-g_s^2|^2\,ds,
\end{align}
where the second inequality follows from \Cref{lem:Xi_lipschitz}.
\end{proof}

With the stability of the mid-price process established, it remains to
control the corresponding processes $q^{W^{0}}$ and $\varphi$. These estimates,
together with Proposition~\ref{prop:Fstability}, allow us to close the fixed
point argument. Since the solution $P$ of the backward one-dimensional Riccati
equation \eqref{P_eq} is continuous, we define
\begin{equation} \label{cp}
C_P(T):=\sup_{t\in[0,T]}P_t.
\end{equation}
Moreover, by continuity and the terminal condition $P_T=2\xi$,
we have $C_P(T)\to 2\xi$ as $T\to0$.
\begin{theorem}[Contraction Condition]
\label{thm:fixedpoint}
Suppose that \Cref{asmfg} and the condition \eqref{bound1} hold.
Then the mapping $\Psi:\mathcal M_{W^0}\to\mathcal M_{W^0}$, defined in
\eqref{fixedpointmap} and characterized by \eqref{trsystem}, has a unique fixed point, provided that
\begin{equation}
\frac{\sqrt{T C_F(T)}}{2a}
\left(1+\frac{C_PT}{2a}\right)
\left[
1+
\sqrt{
2\left(
1+\frac{C_P^2T^2}{4a^2}
\right)
}
\right]
<1,
\label{mfgcontraction}
\end{equation}
where $C_F(T)$ is given in \Cref{prop:Fstability} and $C_P(T)$ is defined in \eqref{cp}.
\end{theorem}

\begin{proof}
Let $g^1,g^2\in\mathcal M_{W^0}$, and denote the corresponding processes by
$F^i$, $q^{W,i}$, and $\varphi^i$, $i=1,2$.

By Proposition~\ref{prop:Fstability},
\begin{equation}
\mathbb E
\left|
F_t^1-F_t^2
\right|^2
\leq
C_F(T)\,
\mathbb E\int_0^T
|g_s^1-g_s^2|^2\,ds,
\qquad
t\in[0,T].
\label{Ffixedestimate}
\end{equation}
Consequently,
\begin{equation}
\mathbb E\int_0^T
\left|
F_t^1-F_t^2
\right|^2dt
\leq
T C_F(T)
\mathbb E\int_0^T
|g_t^1-g_t^2|^2dt.
\label{Ffixednorm}
\end{equation}

We first propagate this estimate to the backward component $\varphi$.
The difference of the two backward equations satisfies
\begin{equation}
d\left(
\varphi_t^1-\varphi_t^2
\right)
=
\frac{P_t}{2a}
\left(
F_t^1-F_t^2
+
\varphi_t^1-\varphi_t^2
\right)dt
+
\left(
Z_t^1-Z_t^2
\right)dW_t^0,
\label{deltaphi}
\end{equation}
with terminal condition
\begin{equation}
\varphi_T^1-\varphi_T^2
=
-\left(
F_T^1-F_T^2
\right).
\end{equation}
Then
\begin{align}
\varphi_t^1-\varphi_t^2
=
-\mathbb E\Bigg[
&
\exp\left(
-\int_t^T\frac{P_r}{2a}\,dr
\right)
\left(
F_T^1-F_T^2
\right)
+
\int_t^T
\exp\left(
-\int_t^s\frac{P_r}{2a}\,dr
\right)
\frac{P_s}{2a}
\left(
F_s^1-F_s^2
\right)ds
\ \Bigg|\ \mathcal F_t^{W^{0}}
\Bigg].
\label{phirepresentation}
\end{align}
Since $P_t\geq0$, conditional Jensen's inequality and the
Cauchy--Schwarz inequality yield
\begin{align}
\mathbb E
\left|
\varphi_t^1-\varphi_t^2
\right|^2
&\leq
2\mathbb E
\left|
F_T^1-F_T^2
\right|^2
+
\frac{C_P^2}{2a^2}
(T-t)
\int_t^T
\mathbb E
\left|
F_s^1-F_s^2
\right|^2ds
\nonumber\\
&\leq
2\left(
1+\frac{C_P^2T^2}{4a^2}
\right)
C_F(T)
\mathbb E\int_0^T
|g_s^1-g_s^2|^2ds.
\label{phipointestimate}
\end{align}
It follows that
\begin{equation}
\mathbb E\int_0^T
\left|
\varphi_t^1-\varphi_t^2
\right|^2dt
\leq
2T
\left(
1+\frac{C_P^2T^2}{4a^2}
\right)
C_F(T)
\mathbb E\int_0^T
|g_t^1-g_t^2|^2dt.
\label{phifixednorm}
\end{equation}

We next estimate the corresponding difference of the conditional mean
inventories. From the forward equation,
\begin{align}
d\left(
q_t^{W_0,1}-q_t^{W_0,2}
\right)
&=
-\frac{P_t}{2a}
\left(
q_t^{W_0,1}-q_t^{W_0,2}
\right)dt
-\frac{1}{2a}
\left(
F_t^1-F_t^2
+
\varphi_t^1-\varphi_t^2
\right)dt,
\label{deltaqW}
\end{align}
with $q_0^{W,1}-q_0^{W,2}=0$. Therefore,
\begin{align}
q_t^{W_0,1}-q_t^{W_0,2}
=
-\frac{1}{2a}
\int_0^t
&
\exp\left(
-\int_s^t\frac{P_r}{2a}\,dr
\right)
\left(
F_s^1-F_s^2
+
\varphi_s^1-\varphi_s^2
\right)ds.
\end{align}
Since $P_t\geq0$, the Cauchy--Schwarz inequality gives
\begin{align}
\left(
\mathbb E\int_0^T
\left|
q_t^{W_0,1}-q_t^{W_0,2}
\right|^2dt
\right)^{1/2}
&\leq
\frac{T}{2a}
\Bigg[
\left(
\mathbb E\int_0^T
\left|
F_t^1-F_t^2
\right|^2dt
\right)^{1/2}
+
\left(
\mathbb E\int_0^T
\left|
\varphi_t^1-\varphi_t^2
\right|^2dt
\right)^{1/2}
\Bigg].
\label{qWfixednorm}
\end{align}

Finally, by \eqref{fixedpointmap},
\begin{align}
\Psi(g^1)_t-\Psi(g^2)_t
=
-\frac{1}{2a}
\Big(
&
F_t^1-F_t^2
+
P_t
\left(
q_t^{W_0,1}-q_t^{W_0,2}
\right)
+
\varphi_t^1-\varphi_t^2
\Big).
\end{align}
Hence,
\begin{align}
\|\Psi(g^1)-\Psi(g^2)\|_{\mathcal M_{W^0}}
&\leq
\frac{1}{2a}
\left(1+\frac{C_PT}{2a}\right)
\Bigg[
\left(
\mathbb E\int_0^T
\left|
F_t^1-F_t^2
\right|^2dt
\right)^{1/2}
+
\left(
\mathbb E\int_0^T
\left|
\varphi_t^1-\varphi_t^2
\right|^2dt
\right)^{1/2}
\Bigg]
\nonumber\\
&\leq
\frac{\sqrt{T C_F(T)}}{2a}
\left(1+\frac{C_PT}{2a}\right)
\left[
1+
\sqrt{
2\left(
1+\frac{C_P^2T^2}{4a^2}
\right)
}
\right]
\|g^1-g^2\|_{\mathcal M_{W^0}}.
\end{align}
Thus, under \eqref{mfgcontraction}, $\Psi$ is a contraction on
$\mathcal M_{W^0}$. The Banach fixed point theorem yields a unique fixed point.
\end{proof}
If the contraction condition \eqref{mfgcontraction} holds, we denote the unique fixed point of $\Psi$ by $\bar{v}$.
Finally, we note that the contraction condition \eqref{mfgcontraction} is
automatically satisfied for a sufficiently small time horizon. Indeed, by
the definition of $C_F(T)$ and the estimates above,
\begin{equation}
C_F(T)\longrightarrow0
\qquad\text{as }T\longrightarrow0.
\end{equation}
Consequently, the left-hand side of \eqref{mfgcontraction} converges to zero
as $T\to0$. Hence, the  mean field game system described by \eqref{trsystem} admits a unique solution
for all sufficiently small $T>0$.  We conclude this section by summarizing the above results in the following theorem.
\begin{theorem}[Fixed Point Characterization and Nash Equilibrium]
Suppose that \Cref{asmfg} and the contraction condition
\eqref{mfgcontraction} hold. Then there exists a unique fixed-point solution
\begin{equation} \label{somfg}
\left(
\bar u^a,\bar u^b,\bar F,\bar q^{W^{0}},\bar\varphi,\bar v
\right)
\end{equation}
to the mean field consistency condition characterized by \eqref{trsystem}. Moreover, the optimal feedback strategy
$v_t^{*}$ given by
\begin{equation} \label{eqst}
v_t^{*}
=
-\frac{1}{2a}
\left(
\bar F_t+P_t(q_t^{*}-r)+\bar \varphi_t
\right),\qquad t\in[0,T],
\end{equation}
where $P_t$ is the solution of \eqref{P_eq} and $q^*_t$ satisfies 
\begin{equation}\label{q_optimal_equilibrium}
dq_t^*
=
-\frac{P_t}{2a}(q_t^{*}-r)\,dt
-\frac{\bar{F}_t+\bar{\varphi}_t}{2a}\,dt
+\sigma^q\,d\beta_t,\qquad q^*_0\in \mathbb{R}_{>0},
\end{equation}
forms a Nash equilibrium
for the limiting system \eqref{limsys}-\eqref{limcost}.
\end{theorem}
\begin{proof}
The representative trader's optimal control problem, for a given mean field
$\bar v$, is solved in Section~\ref{sincon}. In particular, the feedback law
$v^{i,*}$ minimizes $J_i^\infty(v^i;\bar v)$, that is,
\begin{equation}
J_i^\infty(v^{i,*};\bar v)
=
\inf_{v^i\in\mathcal M_{\mathcal F^{W^{0},\beta^i}}}
J_i^\infty(v^i;\bar v).
\end{equation}
Moreover, the conditional mean inventory is given by
\begin{equation}
\bar q_t^{W^{0}}
=
\mathbb E\left[
q_t^{i,*}\mid\mathcal F_t^{W^{0}}
\right]
=
\mathbb E[q_0]
-
\int_0^t
\frac{
P_s(\bar q_s^{W^{0}}-r)+\bar F_s+\bar\varphi_s
}{
2a
}\,ds. 
\end{equation}
taking the conditional expectation of the optimal feedback law yields
\begin{align}
\mathbb E\left[
v_t^{i,*}\mid\mathcal F_t^{W^{0}}
\right]
&=
\mathbb E\left[
-\frac{
\bar F_t+P_t(q_t^{i,*}-r)+\bar\varphi_t
}{
2a}
\bigg|\mathcal F_t^{W^{0}}
\right]
\nonumber\\
&=
-\frac{
\bar F_t+P_t(\bar q_t^{W^{0}}-r)+\bar\varphi_t
}{
2a}.
\end{align}
Since
$
\left(
\bar u^a,\bar u^b,\bar F,\bar q^{W^{0}},\bar\varphi,\bar v
\right)
$
is the solution of \eqref{trsystem}, the last equation in \eqref{trsystem}
yields
\begin{equation}
-\frac{
\bar F_t+P_t(\bar q_t^{W^{0}}-r)+\bar\varphi_t
}{
2a}
=
\bar v_t.
\end{equation}
Therefore,
$
\mathbb E\left[
v_t^{i,*}\mid\mathcal F_t^{W^{0}}
\right]
=
\bar v_t,
$
and hence the consistency condition is satisfied.
\end{proof}
\subsection{Numerical Experiments: A Special Case}\label{sec:simulation}
In this section, we provide a numerical illustration of 
the mean field equilibrium system \eqref{trsystem} in a particular case.
As in Section~4 of \cite{cont2021}, we consider a setting in which
the density-price system $\mathrm{UF}[g]$, described by \eqref{loblim-1}--\eqref{trf}, reduces to a scalar system. 

In short, suppose that the drift terms in the limit order book density
dynamics are of the form 
\begin{equation}
G^l(g,u,t)+J^l(u,t)
=
c^l(g)h_1^{l,*},
\qquad
l\in\{a,b\},
\end{equation}
where $h_1^{l,*}$ denotes the first eigenfunction of $A_l^*$, i.e.,
$A_l^*h_1^{l,*}=\lambda_1^l h_1^{l,*}$ and
$\|h_1^{l,*}\|_{H^l}=1$, for $l\in\{a,b\}$ (see also \eqref{eignh}), and
$c^l:\mathbb R\to\mathbb R$ is a measurable scalar-valued function. The
initial condition is given by $u_0^l=d_0^l h_1^{l,*},
 d_0^l>0$. The functions $c^a$ and $c^b$ are strictly positive, with $c^a$
nonincreasing and $c^b$ nondecreasing. This specification is consistent
with the economic interpretation of the model, as aggregate buying has a
negative effect on the ask-side depth, while aggregate selling has a
negative effect on the bid-side depth. Here, $W^a$ and $W^b$ are one-dimensional independent Wiener processes.

Under these assumptions, the market depth dynamics $D^a,D^b$ can be reduced
to standard one-dimensional diffusion processes without involving the
density processes $u^a,u^b$; we refer to Section~4 of \cite{cont2021} for
further details on this reduction. Moreover, as discussed in \Cref{sec:truncated_density_price}, the truncated
and untruncated versions of the mid-price dynamics are arbitrarily close for
sufficiently large upper and sufficiently small lower truncation bounds, and
the numerical results are barely affected by whether such truncation is
imposed. Therefore, in this section only, for simplicity and to reduce the
number of parameters, we use the untruncated version of the mid-price
dynamics. Consequently, the mean field equilibrium system \eqref{trsystem}
reduces to the nonlinear FBSDE system given by
\begin{equation}
\begin{cases}
d\bar D_t^a
=
\lambda_a
\big(
c^a(\bar v_t)-\bar D_t^a
\big)\,dt
+
\sigma_a \bar D_t^a\,dW_t^a,
\qquad \bar D_0^a\in \mathbb{R}_{>0},
\\[2mm]
d\bar D_t^b
=
\lambda_b
\big(
c^b(\bar v_t)-\bar D_t^b
\big)\,dt
+
\sigma_b \bar D_t^b\,dW_t^b,
\qquad \bar D_0^b \in \mathbb{R}_{>0},
\\[2mm]
d\bar F_t
=
\kappa\left[
\frac{
\lambda_b\big(c^b(\bar v_t)-\bar D_t^b\big)
}{
\bar D_t^b
}
-
\frac{
\lambda_a\big(c^a(\bar v_t)-\bar D_t^a\big)
}{
\bar D_t^a
}
\right]dt
+
\kappa\sigma_b\,dW_t^b
-
\kappa\sigma_a\,dW_t^a,
\qquad \bar F_0 \in \mathbb{R}_{>0},
\\[3mm]
d\bar q^{W^{0}}_t
=
\bar v_t\,dt,
\qquad
\bar q_0^{W^0}=\mathbb{E}[\bar{q}_0],
\\[2mm]
d\bar p_t
=
-2\phi\bar q^{W^{0}}_t\,dt
+
\bar z_t^a\,dW_t^a
+
\bar z_t^b\,dW_t^b,
\qquad
\bar p_T
=
2\xi\bar q_T-\bar F_T,
\\[2mm]
\bar v_t
=
-\dfrac{\bar F_t+\bar p_t}{2a},
\end{cases}
\label{eq:numerical_mfg}
\end{equation}
where $\lambda_a,\lambda_b,\sigma_a,\sigma_b,\kappa,\gamma,\alpha,$ and $a$
are positive constants. The adjoint process admits the affine representation 
\begin{equation}
\bar p_t=P_t\bar q^{W^{0}}_t+\bar\varphi_t,
\end{equation}
where $P$ and $\bar\varphi$ satisfy
\begin{equation}
\begin{cases}
\displaystyle
\dot P_t
=
\frac{P_t^2}{2a}
-
2\phi,
\qquad
P_T=2\xi,
\\[2mm]
\displaystyle
d\bar\varphi_t
=
\frac{P_t}{2a}
\left(
F_t+\bar\varphi_t
\right)dt
+
\bar Z_t^a\,dW_t^a
+
\bar Z_t^b\,dW_t^b,
\qquad
\bar\varphi_T=-\bar F_T.
\end{cases}
\end{equation}
For the numerical implementation, however, we work directly with the
adjoint process $\bar p$, which reduces the number of equations involved in
the simulation.

To solve \eqref{eq:numerical_mfg} numerically, we employ a Picard iteration
following the same order of construction as in the contraction argument developed in \Cref{sec:fixed_point_equilibrium}. Specifically, we start with an initial guess $g^{(0)}$ for the mean field and iterate according to 
\begin{equation}
g^{(k)}
\longmapsto
\left(D^{a,(k)},D^{b,(k)}\right)
\longmapsto
F^{(k)}
\longmapsto
\left(q^{W^{0},(k)},p^{(k)}\right) 
\longmapsto
g^{(k+1)},
\end{equation}
until converges, which yields $(\bar{D}^a,\bar{D}^b, \bar{F},\bar{q},\bar{p}, \bar{v})$. For the FBSDE component, we use the regression-based Monte Carlo method
in \cite{GobetLemorWarin2005}. In
particular, the conditional expectations arising in the backward recursion
are approximated by least-squares projections onto a finite-dimensional
basis.

For the numerical experiment, we take  
\begin{equation}
c^a(x)
=
c-\left(x^+\wedge d\right),
\qquad
c^b(x)
=
c-\left(x^-\wedge d\right),
\end{equation}
where $x^+=x\vee0, \,x^-=(-x)\vee0,\,c>d>0$.

\begin{figure}[htbp]
    \centering
    \begin{minipage}{0.48\textwidth}
        \centering
        \includegraphics[width=\textwidth]{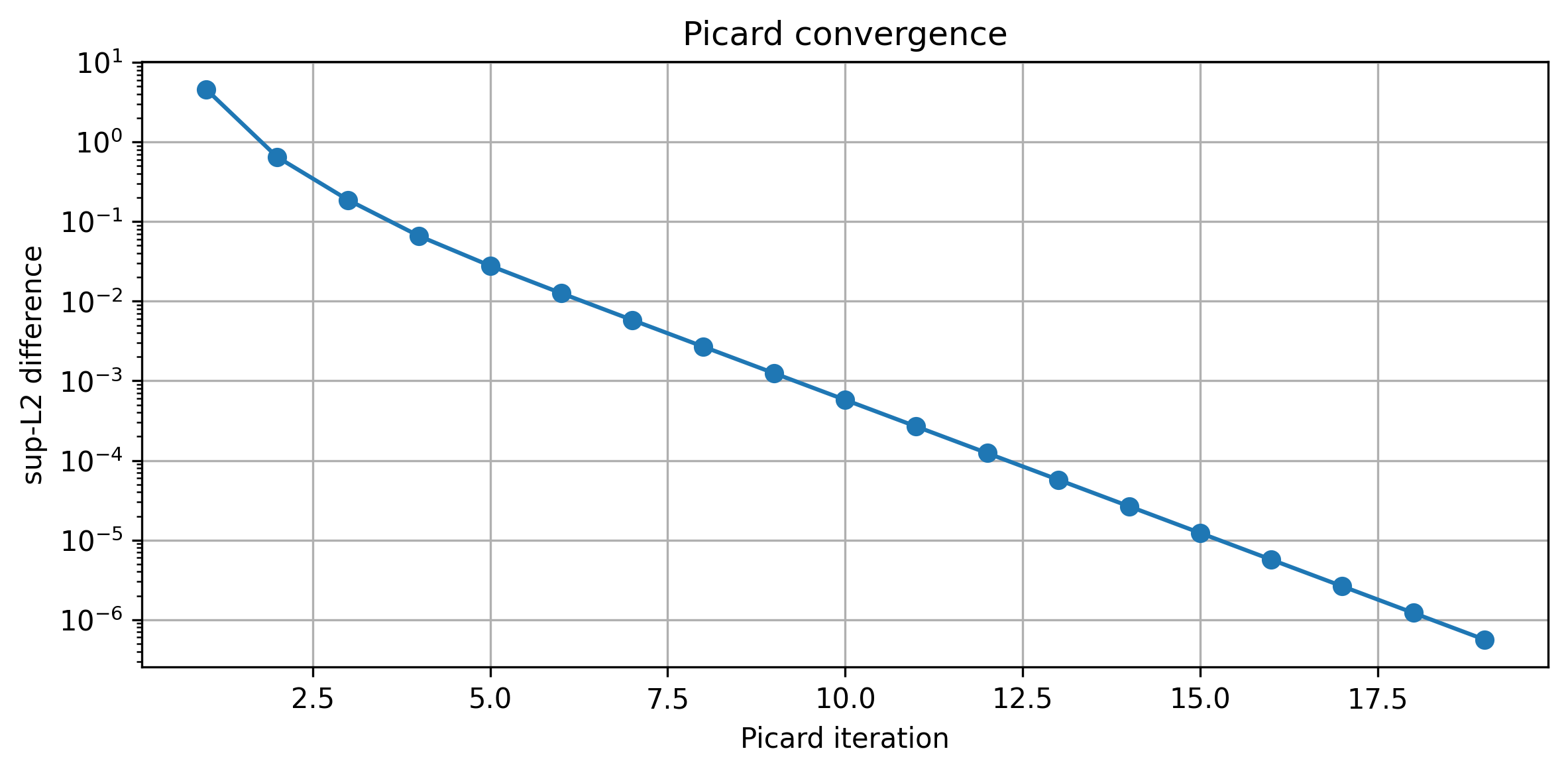}
        \caption*{(a) Picard iteration.}
    \end{minipage}
    \hfill
    \begin{minipage}{0.48\textwidth}
        \centering
        \includegraphics[width=\textwidth]{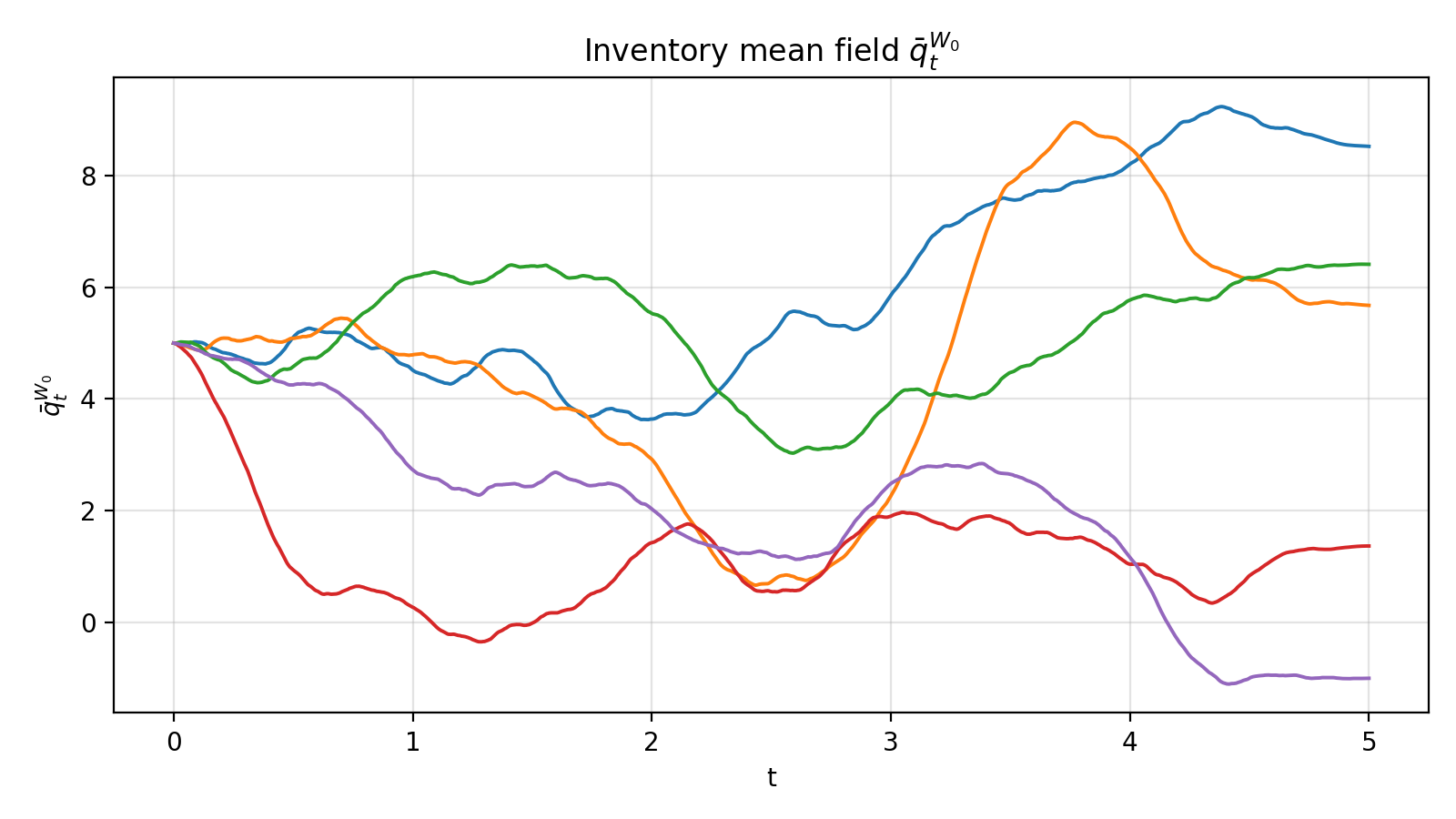}
        \caption*{(b) Inventory mean field $\bar q_t^{W_0}$.}
    \end{minipage}

    \vspace{3mm}

    \begin{minipage}{0.48\textwidth}
        \centering
        \includegraphics[width=\textwidth]{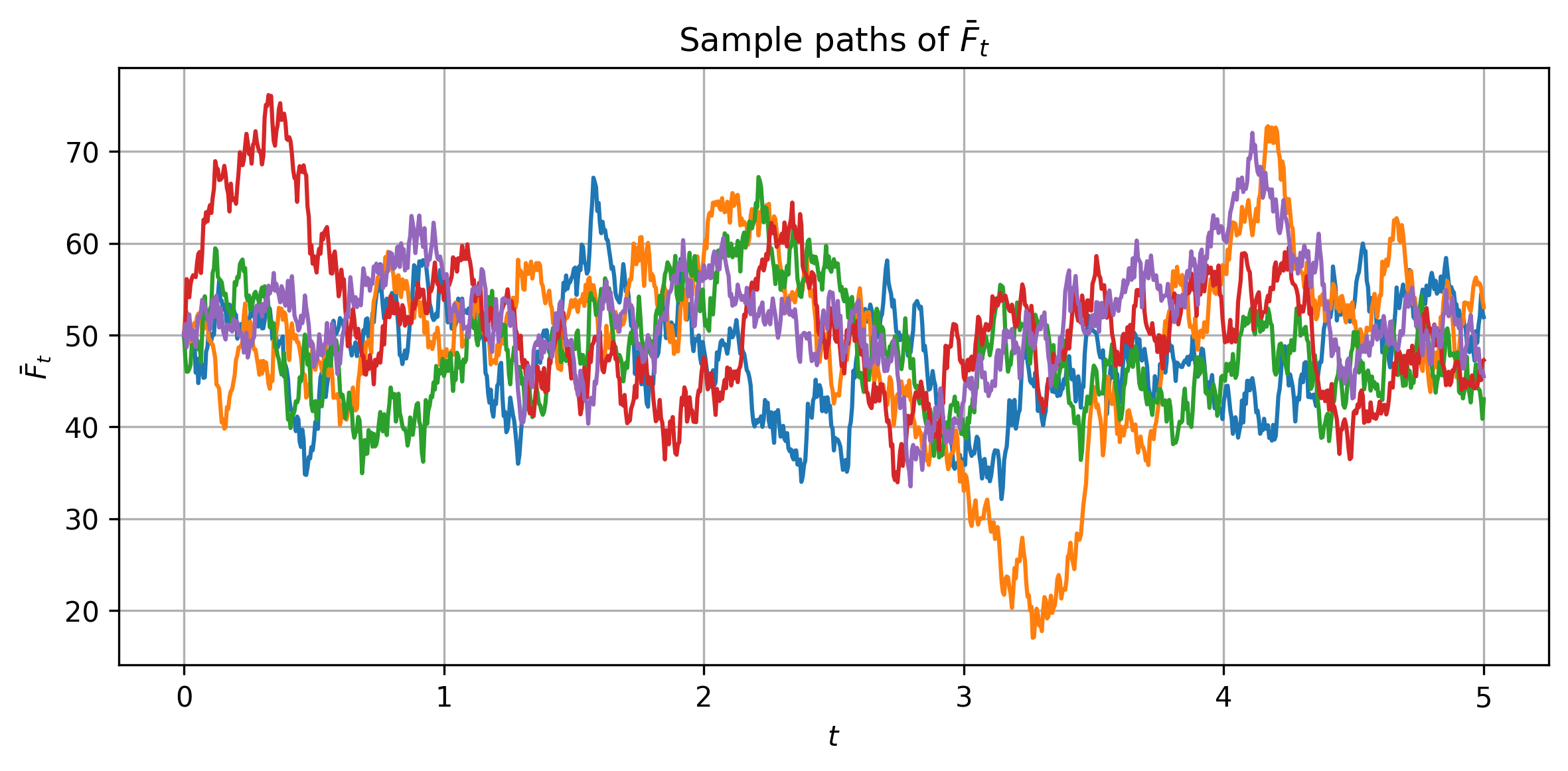}
        \caption*{(c) Equilibrium mid-price $\bar F_t$.}
    \end{minipage}
    \hfill
    \begin{minipage}{0.48\textwidth}
        \centering
        \includegraphics[width=\textwidth]{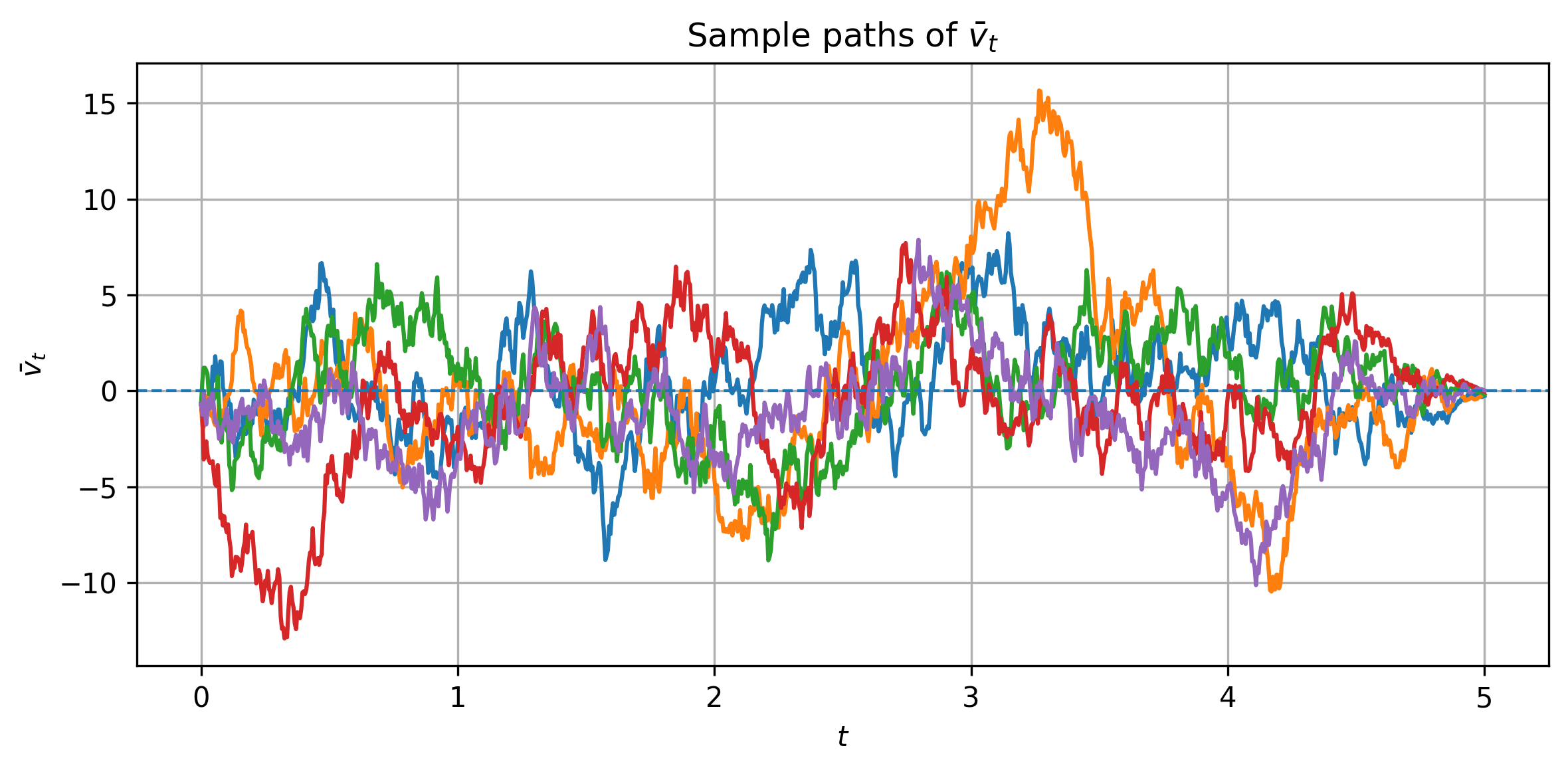}
        \caption*{(d) Trading rate mean field $\bar v_t$.}
    \end{minipage}

    \caption{Numerical results for the special case.}
    \label{fig:numerical_results}
\end{figure}

The functions $c^a$ and $c^b$ represent the mean-reversion levels of the
ask- and bid-side market depths, respectively. In the absence of aggregate
trading, both depths revert toward the baseline level $c$. Aggregate buying
lowers the mean-reversion level of the ask-side depth, as market buy orders
reduce the ask-side depth, while aggregate selling analogously lowers the
mean-reversion level of the bid-side depth. The truncation at $d<c$ ensures
that these mean-reversion levels remain strictly positive.
We use $T=5$, $\bar F_0=50$, $\bar D_0^a=\bar D_0^b=40$, $\lambda_a=\lambda_b=1.5$,
$\sigma_a=\sigma_b=2$, $\kappa=7$, $c=50$, $d=48$, $q_0=5$,
$a=1$, $r=0$, $\phi=0.03$, and $\xi=0.03$. The Picard iteration for $g$ converges with a final error of
$5.64\times10^{-7}$, while the average relative least-squares error for the
backward approximation is approximately $4.4\times10^{-3}$, providing
numerical evidence for the existence of a fixed point and the accuracy of
the backward approximation in this special case. The corresponding simulated sample
paths of the inventory mean field $\bar q$, the mid-price $\bar F$, and the
trading rate mean field $\bar v$ are presented in \Cref{fig:numerical_results}.

\section{$\epsilon$-Nash Equilibrium}
\label{sec:epsilon-nash}
In this section, we establish that the mean field equilibrium strategy
given by \eqref{eqst} yields an $\epsilon$-Nash equilibrium for the
$N$-player game. Here, we use the superscript $[N]$ to indicate processes related to the $N$-player game, and a bar to denote the solutions of the mean field consistency condition given by \eqref{somfg}. We first recall the dynamics and cost functional of the $N$-player game.
The inventory process of trader $i$ satisfies
\begin{equation} \label{Nstate}
dq_t^{[N],i}
=
v_t^{[N],i}\,dt
+
\sigma^q\,d\beta_t^i,
\qquad
q_0^{[N],i}=q_0^i\in \mathbb{R}_{>0},
\qquad i=1,\ldots,N.
\end{equation}
The empirical average trading rate is given by
\begin{equation}
v_t^{(N)}
=
\frac{1}{N}\sum_{i=1}^N v_t^{[N],i}.
\end{equation}
The corresponding density-price system, denoted by
\begin{equation}
UF\left[v^{(N)}\right]
=
\left(
u^{a}\left[v^{(N)}\right],
u^{b}\left[v^{(N)}\right],
F\left[v^{(N)}\right]
\right)
:=
\left(u^{a,[N]},u^{b,[N]},F^{[N]}\right),
\end{equation}
is then given by \eqref{loblim-1}--\eqref{trf} with $g:=v^{(N)}$. The execution price and cash process of trader $i$ are then given by
\begin{align}
S_t^{[N],i}
&=
F_t^{[N]}+av_t^{[N],i}, \qquad S_0^{[N],i}=S_0^i \in \mathbb{R}_{>0},
\\
dZ_t^{[N],i}
&=
-S_t^{[N],i}\,dq_t^{[N],i}, \qquad Z_0^{[N],i}=Z_0^i \in \mathbb{R}_{>0}. \label{Znste}
\end{align}

The cost functional of trader $i$ in the $N$-player game is given by
\begin{equation}\label{costN}
J_i^{[N]}
\left(
v^{[N],i};v^{[N],-i}
\right)
=
\mathbb E
\left[
\int_0^T
\left(
a\left|v_t^{[N],i}\right|^2
+
F_t^{[N]} v_t^{[N],i}
+
\phi\left|q_t^{[N],i}-r\right|^2
\right)dt
+
\xi\left|q_T^{[N],i}-r\right|^2
-
(q_T^{[N],i}-r)F_T^{[N]}
\right].
\end{equation}

We now consider the strategy profile
$\{v^{[N],\circ,i}\}_{i=1}^N$ obtained from the mean field
equilibrium feedback law \eqref{eqst}, where
$v_t^{[N],\circ,i}$ is given by
\begin{equation}\label{control_N_player}
v_t^{[N],\circ,i}
=
-\frac{\bar F_t+P_t(q_t^{[N],\circ,i}-r)+\bar\varphi_t}{2a}.
\end{equation}
In other words, each agent's strategy depends only on her own inventory
process and the common processes $\bar F$ and $\bar\varphi$,
without requiring observation of the states of other agents.

We now recall the definition of an $\epsilon$-Nash equilibrium for the above
$N$-player game. A strategy profile
$\big(v^{[N],\circ,1},\ldots,v^{[N],\circ,N}\big)$
is called an $\epsilon$-Nash equilibrium if, for every trader
$i=1,\ldots,N$,
\begin{equation}
J_i^{[N]}
\left(
v^{[N],\circ,i};v^{[N],\circ,-i}
\right)
\leq
\inf_{v^{[N],i}\in\mathcal M_{\mathcal F^{[N]}}}
J_i^{[N]}
\left(
v^{[N],i};v^{[N],\circ,-i}
\right)
+\epsilon_N,
\label{epsnashdef}
\end{equation}
where $\{\epsilon_N\}_{N\geq1}$ is a nonnegative sequence satisfying
$\epsilon_N\rightarrow0
\,\text{as }N\rightarrow\infty.
$
The above property can equivalently be written as
\begin{equation}
J_i^{[N]}
\left(
v^{[N],\circ,i};v^{[N],\circ,-i}
\right)
\leq
\inf_{v^{[N],i}\in\mathcal A_{\mathcal F^{[N]}}}
J_i^{[N]}
\left(
v^{[N],i};v^{[N],\circ,-i}
\right)
+\epsilon_N,
\end{equation}
where
\begin{equation}
\mathcal A_{\mathcal F^{[N]}} 
:=
\left\{
v^{[N],i}\in\mathcal M_{\mathcal F^{[N]}} 
:
J_i^{[N]}
\left(
v^{[N],i};v^{[N],\circ,-i}
\right)
\leq
J_i^{[N]}
\left(
v^{[N],\circ,i};v^{[N],\circ,-i}
\right)
\right\}.
\end{equation}
Clearly, $\mathcal A_{\mathcal F^{[N]}} $ is nonempty. 
Due to the symmetry among the traders, it suffices to consider the case in
which trader $1$ deviates from the equilibrium strategy and adopts an
alternative admissible control $v^{[N],1}$, while all other traders, indexed by $i=2,\ldots,N$, follow
the control laws given by \eqref{control_N_player}.

Thus, the sequence of $N$-player games, indexed by the number of agents $N$,
is completely determined by the sequence of control actions
$\{v^{[N],1}\}_{N\geq 1}$ adopted by trader $1$. The following lemma will be useful for obtaining a uniform bound for admissible controls.
\begin{lemma} \label{enb1}
Consider the $N$-player game \eqref{Nstate}--\eqref{Znste}. For any admissible
strategy profile $\left(
v^{[N],1},\ldots,v^{[N],N}
\right)$, the strategy of each trader $i=1,\ldots,N$ satisfies
\begin{equation}
\mathbb E\int_0^T
\left|v_t^{[N],i}\right|^2\,dt
\leq
\frac{2}{a}
\left[
J_i^{[N]}
\left(
v^{[N],i};v^{[N],-i}
\right)
+
\left(
\frac{T}{2a}+\frac{1}{4\xi}
\right)C_F
\right],
\end{equation}
where $C_F$ is the bound given in \eqref{boundF}.
\end{lemma}

\begin{proof}
By Young's inequality,
\begin{equation}
F_t^{[N]}v_t^{[N],i}
\ge - |F_t^{[N]}v_t^{[N],i}|
\ge -\frac{a}{2}|v_t^{[N],i}|^2
-\frac{1}{2a}|F_t^{[N]}|^2.
\end{equation}
Hence,
\begin{equation}
a|v_t^{[N],i}|^2+F_t^{[N]}v_t^{[N],i}
\ge
\frac{a}{2}|v_t^{[N],i}|^2
-\frac{1}{2a}|F_t^{[N]}|^2.
\end{equation}
Moreover, since $\xi>0$,
\begin{equation}
\xi(q_T^{[N],i}-r)^2-(q_T^{[N],i}-r)F_T^{[N]}
=
\xi\left(
q_T^{[N],i}-r-\frac{F_T^{[N]}}{2\xi}
\right)^2
-\frac{1}{4\xi}|F_T^{[N]}|^2
\ge
-\frac{1}{4\xi}|F_T^{[N]}|^2.
\end{equation}
Since $\int_0^T |q_t^{[N],i}-r|^2\,dt\ge0$,
it follows that
\begin{align}
J_i^{[N]}
\left(
v^{[N],i};v^{[N],-i}
\right)
&\ge
\frac{a}{2}
\mathbb E\int_0^T |v_t^{[N],i}|^2\,dt
-
\frac{1}{2a}
\mathbb E\int_0^T |F_t^{[N]}|^2\,dt
-
\frac{1}{4\xi}
\mathbb E|F_T^{[N]}|^2.
\end{align}
Using the bound in \eqref{boundF}, we obtain
\begin{equation}
J_i^{[N]}
\left(
v^{[N],i};v^{[N],-i}
\right)
\ge
\frac{a}{2}
\mathbb E\int_0^T |v_t^{[N],i}|^2\,dt
-
\left(
\frac{T}{2a}+\frac{1}{4\xi}
\right)C_F.
\end{equation}
The result follows immediately.
\end{proof}
\begin{proposition}[Uniform $L^2$ bound for admissible strategies]\label{enbond2}
There exists a constant $C_{\mathcal A}>0$ such that
\begin{equation}
\mathbb E\int_0^T
\left|v_t^{[N],i}\right|^2\,dt
\leq C_{\mathcal A}
\end{equation}
holds for all $N\geq 1$, $i=1,\ldots,N$, and
$v^{[N],i}\in\mathcal A_{\mathcal F^{[N]}}$.
\end{proposition}

\begin{proof}
When all traders adopt the control laws derived from the mean field
equilibrium, the corresponding inventory and control processes
$q^{[N],\circ,i}$ and $v^{[N],\circ,i}$ satisfy 
\begin{equation}
\left\{
\begin{aligned}
dq_t^{[N],\circ,i}
&=
v_t^{[N],\circ,i}\,dt
+
\sigma^q\,d\beta_t^i,
\qquad
q_0^{[N],\circ,i}=q_0^i,
\\
v_t^{[N],\circ,i}
&=
-\frac{
\bar F_t
+
P_t(q_t^{[N],\circ,i}-r)
+
\bar\varphi_t
}{
2a
},
\qquad
i=1,\ldots,N.
\end{aligned}
\right.
\end{equation}
Here, $(\bar F,\bar\varphi)$ is given by \eqref{somfg},
which solves \eqref{trsystem}, with
\begin{equation}
d\bar\varphi_t
=
\frac{P_t}{2a}
\left(
\bar F_t+\bar\varphi_t
\right)dt
+
\bar Z_t\,dW^{0}_t,
\qquad
\bar\varphi_T=-\bar F_T.
\end{equation}
Due to the bound \eqref{boundF}, standard estimates yield uniform
second-moment bounds for $\bar\varphi$, $q^{[N],\circ,i}$, and
$v^{[N],\circ,i}$, as well as a uniform bound for the associated cost
\eqref{costN}. We denote the resulting uniform bound by $D_F$. Hence,
\begin{equation}
J_i^{[N]}
\left(
v^{[N],\circ,i};v^{[N],\circ,-i}
\right)
\leq D_F.
\end{equation}
Here, $D_F$ is independent of both $N$ and the trader index $i$. By the definition of $\mathcal A_{\mathcal F^{[N]}} $, for every
$v^{[N],i}\in\mathcal A_{\mathcal F^{[N]}} $,
\begin{equation}
J_i^{[N]}
\left(
v^{[N],i};v^{[N],\circ,-i}
\right)
\leq D_F.
\end{equation}
The result then follows immediately from \Cref{enb1}.
\end{proof}
\begin{proposition}[$L^2$ convergence rate for the average trading rate] \label{boundu}
Let $\left\{v^{[N],1}\right\}_{N\geq 1}$ be any sequence of admissible controls such that
$
v^{[N],1}\in\mathcal A_{\mathcal F^{[N]}} ,
\, N\geq 1$. Then there exists a constant $C>0$, independent of $N$ and the sequence
$\{v^{[N],1}\}_{N\geq1}$, such that
\begin{equation}
\mathbb E\int_0^T
\left|
v_t^{(N)}-\bar v_t
\right|^2dt
\leq
\frac{C}{N}.
\label{eq:mean_control_convergence}
\end{equation}
\end{proposition}

\begin{proof}
Averaging the inventory dynamics over the population and adding and
subtracting $v^{[N],\circ,1}$, we obtain
\begin{align}
q_t^{(N)}
&=
q_0^{(N)}
-
\int_0^t
\frac{
\bar F_s+P_s(q_s^{(N)}-r)+\bar\varphi_s
}{
2a
}\,ds
+
\frac{1}{N}
\int_0^t
\left(
v_s^{[N],1}-v_s^{[N],\circ,1}
\right)ds
+
\frac{\sigma^q}{N}
\sum_{i=1}^N \beta_t^i,
\end{align}
On the other hand,
\begin{equation}
\bar q_t^{W^{0}}
=
\mathbb E[q_0]
-
\int_0^t
\frac{
\bar F_s+P_s(\bar q_s^{W^{0}}-r)+\bar\varphi_s
}{
2a
}\,ds.
\end{equation}
Therefore,
\begin{align}
q_t^{(N)}-\bar q_t^{W^{0}}
&=
\frac{1}{N}
\sum_{i=1}^N
\left(
q_0^i-\mathbb E[q_0]
\right)
-
\int_0^t
\frac{P_s}{2a}
\left(
q_s^{(N)}-\bar q_s^{W^{0}}
\right)ds
+
\frac{1}{N}
\int_0^t
\left(
v_s^{[N],1}-v_s^{[N],\circ,1}
\right)ds
+
\frac{\sigma^q}{N}
\sum_{i=1}^N \beta_t^i.
\end{align}
By standard estimates, the i.i.d.\ assumption on the initial inventories,
and \Cref{enbond2}, we obtain
\begin{equation}
\sup_{t\in[0,T]}
\mathbb E
\left|
q_t^{(N)}-\bar q_t^{W^{0}}
\right|^2
\leq
\frac{C}{N},
\end{equation}
where $C$ is independent of $N$ and of the sequence
$\{v^{[N],1}\}_{N\geq1}$.

Furthermore, by the definition of $v^{(N)}$ and the feedback form of
$v^{[N],\circ,i}$,
\begin{equation}
v_t^{(N)}-\bar v_t
=
\frac{1}{N}
\left(
v_t^{[N],1}-v_t^{[N],\circ,1}
\right)
-
\frac{P_t}{2a}
\left(
q_t^{(N)}-\bar q_t^{W^{0}}
\right).
\end{equation}
Using again \Cref{enbond2}, together with the preceding estimate,
we obtain
\begin{equation}
\mathbb E\int_0^T
\left|
v_t^{(N)}-\bar v_t
\right|^2dt
\leq
\frac{C}{N}.
\end{equation}
This proves the result.
\end{proof}
\begin{theorem}[$\epsilon$-Nash Equilibrium]
Suppose that \Cref{asmfg} and the contraction condition \eqref{mfgcontraction} hold.
Then the strategy profile
$\left(
v^{[N],\circ,1},\ldots,v^{[N],\circ,N}
\right)$ forms an $\epsilon_N$-Nash equilibrium for the $N$-player game described by \eqref{Nstate}--\eqref{Znste}, where
$\epsilon_N
=O\left(\frac{1}{\sqrt N}\right)$.
\end{theorem}

\begin{proof}
Let $\{v^{[N],1}\}_{N\geq1}$ be any sequence of admissible controls such that
\begin{equation}
v^{[N],1}\in\mathcal A_{\mathcal F^{[N]}} ,
\qquad N\geq1.
\end{equation}
For each $N$, let $F^{[N]}$ denote the mid-price process associated with the
empirical average trading rate $v^{(N)}$, i.e., $F^{[N]}:=F[v^{(N)}]$.
By \Cref{prop:Fstability} and \Cref{boundu},
\begin{equation}
\sup_{t\in[0,T]}
\mathbb E
\left|
F_t^{[N]}-\bar F_t
\right|^2
\leq
\frac{C}{N},
\end{equation}
where $C$ is independent of $N$ and the choice of the sequence
$\{v^{[N],1}\}_{N\geq1}$.

Recall that 
\begin{equation}
\label{costinf}
J_1^{\infty}
\left(
v^1;\bar F
\right)
=
\mathbb E
\left[
\int_0^T
\left(
a|v_t^1|^2
+
\bar F_t v_t^1
+
\phi|q_t^1-r|^2
\right)dt
+
\xi|q_T^1-r|^2
-
(q_T^1-r)\bar F_T
\right].
\end{equation}
By \eqref{costN} and \eqref{costinf},
\begin{align}
&J_1^{[N]}
\left(
v^{[N],1};v^{[N],\circ,-1}
\right)
-
J_1^\infty
\left(
v^{[N],1};\bar F
\right)
=
\mathbb E\int_0^T
\left(
F_t^{[N]}-\bar F_t
\right)v_t^{[N],1}\,dt
-
\mathbb E\left[
(q_T^{[N],1}-r)
\left(
F_T^{[N]}-\bar F_T
\right)
\right].
\end{align}
Consequently,
\begin{equation}
\left|
J_1^{[N]}
\left(
v^{[N],1};v^{[N],\circ,-1}
\right)
-
J_1^\infty
\left(
v^{[N],1};\bar F
\right)
\right|
\leq
\left|
\mathbb E\int_0^T
\left(
F_t^{[N]}-\bar F_t
\right)v_t^{[N],1}\,dt
\right|
+
\left|
\mathbb E\left[
(q_T^{[N],1}-r)
\left(
F_T^{[N]}-\bar F_T
\right)
\right]
\right|.
\end{equation}
By the Cauchy--Schwarz inequality, the uniform energy bound on
$\mathcal A_{\mathcal F^{[N]}} $, and standard estimates for the inventory
process,
\begin{equation}
\left|
J_1^{[N]}
\left(
v^{[N],1};v^{[N],\circ,-1}
\right)
-
J_1^\infty
\left(
v^{[N],1};\bar F
\right)
\right|
\leq
\frac{C}{\sqrt N}.
\label{costapprox}
\end{equation}
where $C$ is independent of $N$ and of the sequence
$\{v^{[N],1}\}_{N\geq1}$.
The case where all traders use the equilibrium controls is included in the above setting by taking $v^{[N],1}=v^{[N],\circ,1}$. Moreover, by the optimality of the feedback control for the limiting control problem,
\begin{equation}
J_1^\infty
\left(
v^{[N],\circ,1};\bar F
\right)
\leq
J_1^\infty
\left(
v^{[N],1};\bar F
\right).
\end{equation}
Thus, applying the above cost estimate first to
$v^{[N],\circ,1}$ and then to $\forall v^{[N],1}\in\mathcal A_{\mathcal F^{[N]}}  $, we obtain, 
\begin{align}
J_1^{[N]}
\left(
v^{[N],\circ,1};v^{[N],\circ,-1}
\right)
&\leq
J_1^\infty
\left(
v^{[N],\circ,1};\bar F
\right)
+
\frac{C}{\sqrt N}
\nonumber\\
&\leq
J_1^\infty
\left(
v^{[N],1};\bar F
\right)
+
\frac{C}{\sqrt N}
\nonumber\\
&\leq
J_1^{[N]}
\left(
v^{[N],1};v^{[N],\circ,-1}
\right)
+
\frac{2C}{\sqrt N}.
\end{align}
Since the constant $C$ is independent of the choice of
$\{v^{[N],1}\}_{N\geq1}$, taking the infimum over
$v^{[N],1}\in\mathcal A_{\mathcal F^{[N]}} $ yields
\begin{align}
J_1^{[N]}
\left(
v^{[N],\circ,1};v^{[N],\circ,-1}
\right)
&\leq
\inf_{v^{[N],1}\in\mathcal A_{\mathcal F^{[N]}} }
J_1^{[N]}
\left(
v^{[N],1};v^{[N],\circ,-1}
\right)
+
\frac{2C}{\sqrt N}
\nonumber\\
&=
\inf_{v^{[N],1}\in\mathcal M_{\mathcal F^{[N]}}}
J_1^{[N]}                 
\left(
v^{[N],1};v^{[N],\circ,-1}
\right)
+
\frac{2C}{\sqrt N}.
\end{align}
Recalling the equivalent formulation of \eqref{epsnashdef}, we obtain the
$\epsilon_N$-Nash property with $\epsilon_N=O\left(N^{-1/2}\right)$. By symmetry, the same conclusion holds for every trader
$i=1,\ldots,N$.
\end{proof}
\section*{Acknowledgements} The authors would like to thank Rama Cont for helpful comments on an earlier version of this paper during his visit to the University of Toronto in October 2026. 

\section*{Funding} This work was supported by the Natural Sciences and Engineering Research Council of Canada (NSERC) under grants RGPIN-2022-05337 and DGECR-2022-00468.
\section*{Data Availability Statement} No empirical data were used in this study. The numerical results were generated by simulation, using the model and parameters specified in \Cref{sec:simulation}.

\bibliographystyle{elsarticle-num-names} 

\bibliography{sample}

\end{document}